\newif\ifdraft
\newif\iffull
\newif\ifcomment
\newif\iflatexdiff
\newif\ifbibtex
\newif\ifpreprint
\bibtextrue
\preprinttrue
\fulltrue
\def\dvers{v2.3}

\newcommand{\papertitle}{Evolution of the longitudinal and azimuthal  structure \\ of the near-side jet peak in Pb--Pb collisions at \snn\ = \unit[2.76]{TeV}}
\newcommand{\paperabstract}{In two-particle angular correlation measurements, jets give rise to a near-side peak, formed by particles associated to a higher $\pt$ trigger particle. 
Measurements of these correlations as a function of pseudorapidity (\Deta) and azimuthal (\Dphi) differences are used to extract the centrality and $\pt$ dependence of the shape of the near-side peak in the \pt range  $1 <~\pt <$~\unit[8]{\gevc} in Pb--Pb and pp collisions at \snn\ = \unit[2.76]{TeV}. 
A combined fit of the near-side peak and long-range correlations is applied to the data and the peak shape is quantified by the variance of the distributions.
While the width of the peak in the $\Dphi$ direction is almost independent of centrality, a significant broadening in the $\Deta$ direction is found from peripheral to central collisions. This feature is prominent for the low $\pt$ region and vanishes above \unit[4]{\gevc}. The widths measured in peripheral collisions are equal to those in pp in the $\Dphi$ direction and above \unit[3]{\gevc} in the $\Deta$ direction. Furthermore, for the 10\% most central collisions and $1 < \pta <$~\unit[2]{\gevc}, $1 < \ptt <$~\unit[3]{\gevc} a departure from a Gaussian shape is found: a depletion develops around the centre of the peak.
The results are compared to AMPT model simulations as well as other theoretical calculations indicating that the broadening and the development of the depletion is connected to the strength of radial and longitudinal flow.}

\ifpreprint
\documentclass[ALICE,manyauthors]{cernphprep}
\usepackage[comma,square,numbers,sort&compress]{natbib}
\else
\documentclass[10pt,aps,prc,twocolumn,superscriptaddress,altaffilletter,nobibnotes,nofootinbib]{revtex4-1}
\usepackage{preprintcover}
\PreprintIdNumber{CERN-EP-2016-228}
\PreprintCoverPaperTitle{\papertitle}
\PreprintCoverAbstract{\paperabstract}
\PreprintJournalName{PRC}
\fi
\usepackage{hyperref}
\usepackage{graphicx}  
\usepackage{dcolumn}   
\usepackage{bm}        
\usepackage{amssymb}   
\usepackage{amsfonts}
\usepackage{graphics}
\usepackage{epsfig}
\usepackage[usenames]{color}
\usepackage{multirow}
\usepackage{units}
\usepackage{epstopdf}
\RequirePackage{subfigure}
\iflatexdiff
\RequirePackage[normalem]{ulem}
\RequirePackage{color}\definecolor{RED}{rgb}{1,0,0}\definecolor{BLUE}{rgb}{0,0,1}

\fi
\ifpreprint
\else

\fi

\newcommand{\gevc}         {GeV/\ensuremath{c}}

\newcommand{\mevcc}        {MeV/\ensuremath{c^2}}

\newcommand{\ptt}          {\ensuremath{p_{\mathrm{T, trig}}}}
\newcommand{\pta}          {\ensuremath{p_{\mathrm{T, assoc}}}}

\newcommand{\ITS}          {\rm{ITS}}

\newcommand{\ZDCs}         {\rm{ZDCs}}
\newcommand{\SPD}          {\rm{SPD}}
\newcommand{\SDD}          {\rm{SDD}}
\newcommand{\SSD}          {\rm{SSD}}
\newcommand{\TPC}          {\rm{TPC}}

\newcommand{\pp}           {pp}

\newcommand{\s}            {\ensuremath{\sqrt{s}}}
\newcommand{\pt}           {\ensuremath{p_{\mathrm{T}}}{ }}

\newcommand{\snn}          {\ensuremath{\sqrt{s_{\mathrm{NN}}}}}

\newcommand{\dd}           {\ensuremath{\mathrm{d}}}

\newcommand{\Dphi}         {\ensuremath{\Delta\varphi}}
\newcommand{\Deta}         {\ensuremath{\Delta\eta}}
\newcommand{\Ntrig}        {\ensuremath{N_{\mathrm{trig}}}}

\newcommand{\dNassoc}      {\ensuremath{\frac{\dd^2N_{\mathrm{assoc}}}{\dd\Deta\dd\Dphi}}}

\newcommand{\Eq}[1]        {Eq.~\ref{#1}}

\newcommand{\red}[1]       {\textcolor{red}{#1}}

\newcommand{\warn}[1]      {{\small\textbf{\red{(!}\footnote{\textbf{\red{(!)}}~#1}\red{)}}}\marginpar{\textbf{\red{---}}}}

\newcommand{\com}[1]       {}

\iflatexdiff

\renewcommand{\xout}[1]    {\textcolor{red}{\sout{#1}}}
 
\else

\newcommand{\xout}[1]    {}
\fi

\ifdraft
\usepackage{lineno}
\linenumbers
\modulolinenumbers[5]
\usepackage{fancyhdr}
\else
\renewcommand{\warn}[1]{}

\fi
\begin{document}
\newlength{\figlen}
\setlength{\figlen}{\linewidth}
\ifpreprint
\setlength{\figlen}{0.75\textwidth}
\begin{titlepage}
\PHyear{2016}
\PHnumber{228}                   
\PHdate{19 Sep}                  
\title{\papertitle}
\ShortTitle{\papertitle}
\Collaboration{ALICE Collaboration%
         \thanks{See Appendix~\ref{app:collab} for the list of collaboration members}}
\ShortAuthor{ALICE Collaboration} 
\ifdraft
\begin{center}
\today\\ \color{red}DRAFT \dvers\ \hspace{0.3cm} \$Revision: 3200 $\color{white}:$\$\color{black}\vspace{0.3cm}
\end{center}
\fi
\else
\title{\papertitle}
\iffull
\input{authors-prl.tex}
\else
\collaboration{ALICE Collaboration}
\fi
\vspace{0.3cm}
\ifdraft
\date{\today, \color{red}DRAFT \dvers\ \$Revision: 3200 $\color{white}:$\$\color{black}}
\else
\date{\today}
\fi
\fi
\begin{abstract}
\paperabstract
\end{abstract}
\ifpreprint
\end{titlepage}
\setcounter{page}{2}
\else
\maketitle
\fi
\ifdraft
\thispagestyle{fancyplain}
\fi

\section{Introduction}
\label{sec:intro}

In elementary interactions with large
momentum transfer ($Q^2 \gg \Lambda^2_{\rm QCD}$), partons with high transverse momentum ($\pt$) are produced. Carrying net colour charge, they cannot exist freely and, instead, evolve from high to low virtuality producing parton showers. These eventually hadronize into a spray of collimated
hadrons called jets. High-\pt partons are produced at the early stages of heavy-ion collisions. They propagate and evolve through the dense and hot medium  created in these
collisions and are expected to lose energy due to induced gluon radiation and elastic scatterings,
a process commonly referred to as jet quenching. The transfer of energy from the leading parton to
the medium and/or into additional gluon radiation leads to effects that can be
exploited to characterise the colour density and scattering power of the medium.

Experimental methods to study high-\pt parton production differ in their capability to reconstruct the original parton momentum and to characterize the angular and momentum distribution of jet fragments. Furthermore, their sensitivity to experimental bias, most particularly the bias associated with the large underlying-event background encountered in heavy-ion collisions, is different.
Inclusive hadron spectra are unbiased observables, mainly sensitive to
the hadronic fragments with the largest momentum fraction (leading particles). Partonic energy
loss suppresses high-$\pt$ particle yields relative to their production in more
elementary pp and p--A collisions which was observed at RHIC and LHC energies. The largest suppression is observed in central Pb--Pb collisions at the LHC at
$\pt \approx 7$~GeV/$c$ \cite{Abelev:2012hxa, CMS:2012aa}.

Jet reconstruction algorithms have the objective to recombine a maximum of jet fragments within a certain area in the pseudorapidity ($\eta$) - azimuth ($\varphi$) plane in order to obtain the original parton energy and direction. In heavy-ion collisions, due to the large fluctuating energy from particles uncorrelated to the jets, the underlying event, jet reconstruction is limited to high transverse energy and small areas (cone-size) around the parton direction. An inclusive jet suppression commensurable to that of hadrons has been observed at the LHC
\cite{Aad:2012vca, Abelev:2013kqa, Adam:2015ewa} 
together with a large di-jet energy asymmetry \cite{Aad:2010bu, Chatrchyan:2011sx}, suggesting that a large fraction of the lost energy is radiated outside the typical jet cone sizes of $R =$~0.3--0.5. Detailed studies of the energy balance in events with high-energy jets show that the lost energy reappears primarily at low to intermediate $\pt$ (\unit[0.5--3]{GeV/$c$}) outside the jet cone \cite{Chatrchyan:2011sx}. Studies of the momentum and angular distributions of jet fragments show that the jet core is almost unmodified
\cite{Chatrchyan:2012gw, Chatrchyan:2014ava, Aad:2014wha}.

Di-hadron angular correlations represent a powerful complementary tool to study jet quenching and the redistribution of energy in an energy region where jets cannot be identified event-by-event over the fluctuating background and where quenching effects are expected to be large.
Such studies involve measuring the distributions of the relative azimuthal angle $\Delta \varphi$ and pseudorapidity $\Delta \eta$ between particle pairs. The pairs consist of a trigger particle in a certain transverse momentum
$p_{\rm T ,trig}$ interval and an associated particle in a $p_{\rm T,assoc}$ interval.
In these correlations, jets manifest themselves as a peak centred around $(\Delta \varphi = 0,~\Delta \eta = 0 )$ (near-side peak) and a structure elongated in $\Delta\eta$ at $\Delta\varphi = \pi$ (the away-side or recoil-region). At low $\pt$, resonance decays as well as femtoscopic correlations also contribute to the near-side peak. The advantage of using di-hadron correlations is that an event-averaged subtraction of the background from particles uncorrelated to the jet can be performed.
This advantage is shared with the analysis of hadron-jet correlations recently reported in Ref.~\cite{Adam:2015doa,Khachatryan:2016erx}. 

At RHIC, the near-side associated particle yield and peak shape have been studied for different systems and collision energies \cite{Abelev:2009af, Agakishiev:2011st, Adamczyk:2013jei}. Small modifications of the yields with respect to a pp reference from PYTHIA are observed and there is remarkably little dependence on the collision system at the centre-of-mass energies of $\sqrt{s_{\rm NN}} = 62.4$ and $200$~GeV.
An exception is the measurement in central Au--Au collisions at \mbox{$\sqrt{s_{\rm NN}} = 200$~GeV} where the jet-like correlation is substantially broader and the momentum spectrum softer
than in peripheral collisions and than those in collisions of other systems in this kinematic regime. In Ref.~\cite{Agakishiev:2011st}, the broadening observed in central Au--Au collisions at \mbox{$\sqrt{s_{\rm NN}} = 200$~GeV} is seen as an indication of a modified jet fragmentation function.

At the LHC, the measurement of the yield of particles associated to a high-\pt trigger particle (\unit[8--15]{GeV/$c$}) in central Pb--Pb collisions relative to the pp reference at $p_{\rm T, assoc}>3$~GeV/$c$ shows a suppression on the away-side and a moderate enhancement on the near-side indicating that medium-induced modifications can also be expected on the near side \cite{Aamodt:2011vg}. Much stronger modifications are observed for
 lower trigger and associated particle \pt\
($3 < p_{\rm T, trig} < 3.5$~GeV/$c$ and $1 < p_{\rm T, assoc} < 1.5$~GeV/$c$) \cite{Chatrchyan:2012wg,Adam:2016xbp}. In the most central Pb--Pb collisions, the near-side yield is enhanced by a factor of 1.7.

The present paper expands these studies at the LHC to the characterisation of the angular distribution of the associated particles with respect to the trigger particle. The angular distribution is sensitive to the broadening of the jet due to the degradation of its energy and the distribution of radiated energy.
Moreover, possible interactions of the parton shower with the collective longitudinal expansion~\cite{Armesto:2004pt, Armesto:2004vz, Romatschke:2006bb} or with turbulent colour fields~\cite{Majumder:2006wi} in the medium would result in near-side peak shapes that are broader in the $\Delta \eta$ than in the $\Delta \varphi$  direction.
Results from the study of the near-side peak shape of charged particles as a function of centrality and for different combinations of trigger and associated particle \pt are discussed.

The paper is organised in the following way:
the ALICE sub-systems used in the analysis are described in Section~\ref{sec:setup} and the data samples, event and track selection in Section~\ref{sec:selection}. Section \ref{sec:twopartfunc} describes the analysis methods and the systematic uncertainties are discussed in Section~\ref{sec:systunc}.
Results are presented in Section~\ref{sec:results} and conclusions are drawn in Section~\ref{sec:summary}. The key results of the presented analysis are also reported in a short companion paper~\cite{Adam:2016tsv}.

\section{Experimental setup}
\label{sec:setup}

A detailed description of the ALICE detector can be found in Ref.~\cite{Aamodt:2008zz}.
The main subsystems used in the present analysis are the Inner Tracking System~(\ITS) and the
Time Projection Chamber~(\TPC). These have a common acceptance of
$|\eta| < 0.9$ and are operated inside a solenoidal magnetic field of \unit[0.5]{T}.
The \ITS\ consists of six layers of silicon detectors for vertex finding and tracking. The two outermost layers of the \ITS\ constitute of the Silicon Strip Detectors (\SSD), the two middle layers of the Silicon Drift Detectors (\SDD) and the two innermost layers of the Silicon Pixel Detector (\SPD) with the later also used for triggering.
The \TPC\ is the main tracking detector measuring up to 159 space points per track.
The V0 detector, consisting of two arrays of 32 scintillator tiles each, and covering $2.8<\eta<5.1$~(V0-A) and $-3.7<\eta<-1.7$~(V0-C), was used for triggering and
centrality determination~\cite{Aamodt:2010pb,Abelev:2013qoq}. All these detector systems have full azimuthal coverage.

Data from the 2010 and 2011 Pb--Pb runs of the LHC at $\snn=$~\unit[2.76]{TeV} and the 2011 pp run at the same energy are combined in the present analysis. From the 2010 sample, about 16 million minimum-bias Pb--Pb events are considered, while in the 2011 Pb--Pb run about 2 million minimum-bias events and about 21 million centrality-triggered events enhancing the 0--50\% centrality range are used.
The pp event sample consists of 30 million minimum-bias events.

In Pb--Pb collisions, the trigger required a coincidence of signals in both V0-A\ and V0-C. In addition, two Zero Degree Calorimeters~(\ZDCs) for neutron detection located at \unit[$\pm114$]{m} from the interaction point are used to suppress electromagnetic interactions. More details about the event selection can be found in Ref.~\cite{Aamodt:2010cz}. The events are characterized into five collision-centrality classes based on the sum of amplitudes in the V0 detectors~\cite{Abelev:2013qoq} (0--10\% (most central), 10--20\%, 20--30\%, 30--50\% and 50--80\%).
In pp collisions, the trigger required a signal in either of the V0\ detectors or the \SPD~\cite{Abelev:2012sea}.
In both collision systems, these triggers are fully efficient for events entering the two-particle correlation analysis presented in this work.

\section{Event and track selection}
\label{sec:selection}

The collision-vertex position is determined with tracks reconstructed in the \ITS\ and \TPC\ as described
in Ref.~\cite{Abelev:2012hxa}. The vertex reconstruction algorithm is fully efficient for events with at least
one reconstructed primary track within $|\eta|<1.4$~\cite{perfpaper}.
The position of the reconstructed vertex along the beam
direction ($z_{\rm vtx}$) is required to be within \unit[7]{cm} of the detector centre. This value is reduced to \unit[3]{cm} in the study of systematic uncertainties.

The analysis uses tracks reconstructed in the \ITS\ and \TPC\ with $1<\pt<$~\unit[8]{\gevc} and in a fiducial region of $|\eta|<0.8$.
As a first step in the track selection, criteria on the number of space points (at least 70) and the quality of the track fit ($\chi^2/{\rm ndf} < 2$)
in the \TPC\ are applied.
Tracks are further required to have a distance of closest approach to the reconstructed vertex smaller than
\unit[2.4]{cm} and \unit[3.2]{cm} in the transverse and the longitudinal direction, respectively.
Two classes of tracks are combined in order to avoid an azimuthally-dependent tracking efficiency due to inactive \SPD\ modules~\cite{Abelev:2012ej}.
The first class requires for tracks to have at least one hit in the \SPD.
For tracks which do not fulfil this criterion, in the second class, the primary vertex position  is used as additional constraint in the global track fit.
An alternative track selection \cite{ALICE:2011ac}, where a tighter $\pt$-dependent cut on the distance of closest approach to the reconstructed vertex is applied, is used for the assignment of a systematic uncertainty.
Further, the tracks in the second class are required to have a
hit in the first layer of the \SDD.
This modified selection has a less uniform azimuthal acceptance, but includes a smaller number of secondary particles produced by interactions in the detector material or weak decays.

The efficiency and purity of the primary charged-particle selection are estimated from a Monte Carlo~(MC)
simulation using the HIJING~1.383 event generator~\cite{hijing} (for Pb--Pb) and the PYTHIA~6.4 event
generator~\cite{Sjostrand:2006za} with the tune Perugia-0~\cite{Skands:2010ak} (for \pp) with particle transport
through the detector carried out with GEANT3~\cite{geant3ref2}.
The combined efficiency and acceptance of the track reconstruction in $|\eta|<0.8$ is
about 82--85\% at $\pt=$~\unit[1]{\gevc} and decreases to about 76--80\% at $\pt=$~\unit[8]{\gevc} depending on collision system, data sample and event centrality.
The contamination from secondary particles resulting from weak decays and due to interactions in the detector material
decreases from 2.5--4.5\% to 0.5--1\% in the $\pt$ range from 1 to \unit[8]{\gevc}.
The contribution from fake tracks, arising from improperly associated hits, is negligible.
The alternative track selection (see above), has 3--6\% lower combined efficiency and acceptance and about two thirds of the secondary contamination.

Due to the combination of different event samples, see Sec.~\ref{sec:setup}, the number of accepted events per centrality class is not uniform, as is shown in Table~\ref{tab:events}.

\begin{table}[b!] \centering
  \begin{tabular}{ccc}
    \hline
    Collision system & Centrality class & Accepted events ($10^6$) \\
    \hline
    Pb--Pb 	& \phantom{0}0--10\% 	& $\phantom{0}7.7$\\
		& 10--20\%	& $\phantom{0}2.9$ \\
		& 20--30\% 	& $\phantom{0}2.9$ \\
		& 30--50\% 	& $\phantom{0}5.9$ \\
		& 50--80\% 	& $\phantom{0}3.9$ \\ \hline
    pp 		& --- 		& $24.0$ \\
    \hline
  \end{tabular}
  \caption{\label{tab:events}
    Centrality classes and corresponding number of accepted events in pp and Pb--Pb collisions at \mbox{\snn\ = \unit[2.76]{TeV}} used in this analysis.}
\end{table}

\section{Analysis}
\label{sec:twopartfunc}

The correlation between two charged particles (denoted trigger and associated particle) is measured
as a function of the azimuthal angle difference $\Dphi$ (defined within $-\pi/2$ and $3\pi/2$) and pseudorapidity
difference $\Deta$~\cite{alice_pa_ridge}.
The correlation is expressed in terms of the associated yield per trigger particle for intervals of trigger and associated transverse momentum, $\ptt$ and $\pta$, respectively. The $\pt$ intervals can be different or identical, in which case only pairs of particles with $\pta < \ptt$ are considered to avoid double counting. The per-trigger yield can be measured experimentally if the particle distribution is independent of pseudorapidity~\cite{Ravan:2013lwa} in the following way:
\begin{equation}
\frac{1}{\Ntrig} \dNassoc = \frac{S(\Deta,\Dphi)}{B(\Deta,\Dphi)} \label{pertriggeryield}
\end{equation}
where $\Ntrig$ is the total number of trigger particles in the centrality class and $\ptt$ interval, ranging from 0.18 to 36 per event.
The signal distribution
$S(\Deta,\Dphi) = 1/\Ntrig\ \dd^2N_{\rm same}/\dd\Deta\dd\Dphi$
is the associated yield per trigger particle for particle pairs from the same event.
The background distribution $B(\Deta,\Dphi) = \alpha\ \dd^2N_{\rm mixed}/\dd\Deta\dd\Dphi$
corrects for finite pair acceptance and pair efficiency.
It is constructed by correlating the trigger particles in one event with the associated particles from
other events in the same centrality class and within the same \unit[2]{cm}-wide $z_{\rm vtx}$ interval
(each event is mixed with 5--20 events depending on the number of tracks per event).
The background distribution is scaled by a factor $\alpha$ which is chosen such that $B(0, 0)$ is unity for pairs where both particles travel in approximately the same direction (i.e.\ $\Dphi\approx 0,\ \Deta\approx 0$), and thus the efficiency and acceptance for the two particles are identical by construction.
The yield defined by \Eq{pertriggeryield} is constructed for each $z_{\rm vtx}$ interval to account for differences in pair acceptance and efficiency, depending on the vertex position $z_{\rm vtx}$.
The trigger particles and the pairs are corrected for  single-particle efficiency, described below, before the final per-trigger yield is obtained by calculating the average of the $z_{\rm vtx}$ intervals weighted by $\Ntrig$.

A minimum opening angle of the particle pairs is required for both signal and background to avoid a bias due to the reduced efficiency for pairs with small separation. Pairs are required to have a separation of $|\Delta\varphi^{*}_{\rm min}|>$~\unit[0.02]{rad} or $|\Deta|>0.02$, where $\Delta\varphi^*_{\rm min}$ is the minimal azimuthal distance at the same radius between the two tracks within the active detector volume.
Furthermore, correlations induced by secondary particles from long-lived neutral-particle decays ($\rm K^0_s$ and $\Lambda$) and $\gamma$-conversions are suppressed by cutting on the invariant mass ($m_{\rm inv}$) of the particle pair. Pairs are removed which are likely to stem from a $\gamma$-conversion ($m_{\rm inv} <$~\unit[4]{\mevcc}), a K$^0_s$ decay ($|m_{\rm inv} - m({\rm K}^0_s)| <$~\unit[5]{\mevcc}) or a
$\Lambda$ decay ($|m_{\rm inv} - m(\Lambda)| <$~\unit[5]{\mevcc}).
Weak decays of heavier particles give a negligible contribution.

Each trigger and each associated particle is weighted with a correction factor that accounts for detector acceptance, reconstruction efficiencies and contamination from secondary particles. These corrections are applied as a function of $\eta$, $\pt$, $z_{\rm vtx}$ and event centrality. The shape parameters extracted below are expected to be insensitive to these single-particle corrections which was confirmed in the analysis.

The obtained per-trigger yields as a function of relative angle are integrated over particles produced within $|\eta| < 0.8$. As mentioned above, the method requires that the distribution of sources contributing to the correlation are independent of pseudorapidity, which is approximately the case for the inclusive particle distribution~\cite{Aamodt:2010pb} as well as the anisotropic flow~\cite{Adam:2016ows}. It can be easily shown (analytically or in a toy Monte Carlo), that such a pseudorapidity dependence  results in distortions as a function of $\Deta$ of the per-trigger yields which are independent of $\Dphi$. 
In addition, the finite centrality and $z_{\rm vtx}$ bin width in the event mixing has been found to cause $\Deta$-dependent effects due to the dependence of particle production on centrality and the $z_{\rm vtx}$-dependent detector efficiency, respectively. 
In the data, such distortions in $\Deta$, of the order of $0.1\%$, have been observed. While small, these distortions are still relevant compared to the jet-like peak which is on top of the large combinatorial background. In order to suppress distortions of the peak in the $\Deta$ direction, a correction factor is calculated such that the away side, which is outside of the range studied by this work, becomes independent of $\Deta$.  This correction factor is then applied consistently to all $\Dphi$ bins. The correctness of this procedure is supported by the fact that the goodness of the fit (see following section) is substantially improved.

\begin{figure*}[t!]
\centering
\subfigure[][]{\includegraphics[width=0.45\textwidth]{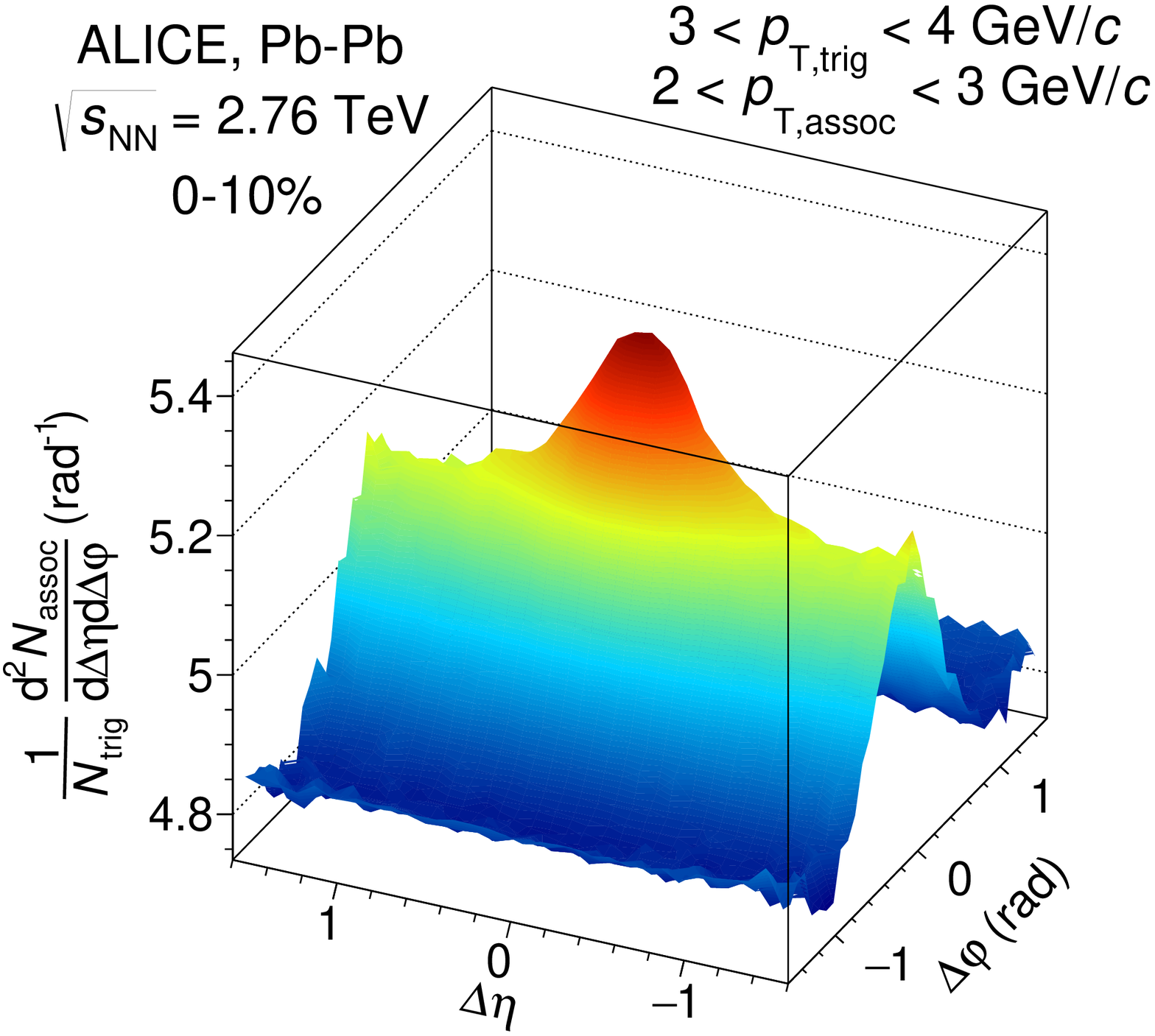} \label{subfig:fit1}}
\subfigure[][]{\includegraphics[width=0.45\textwidth]{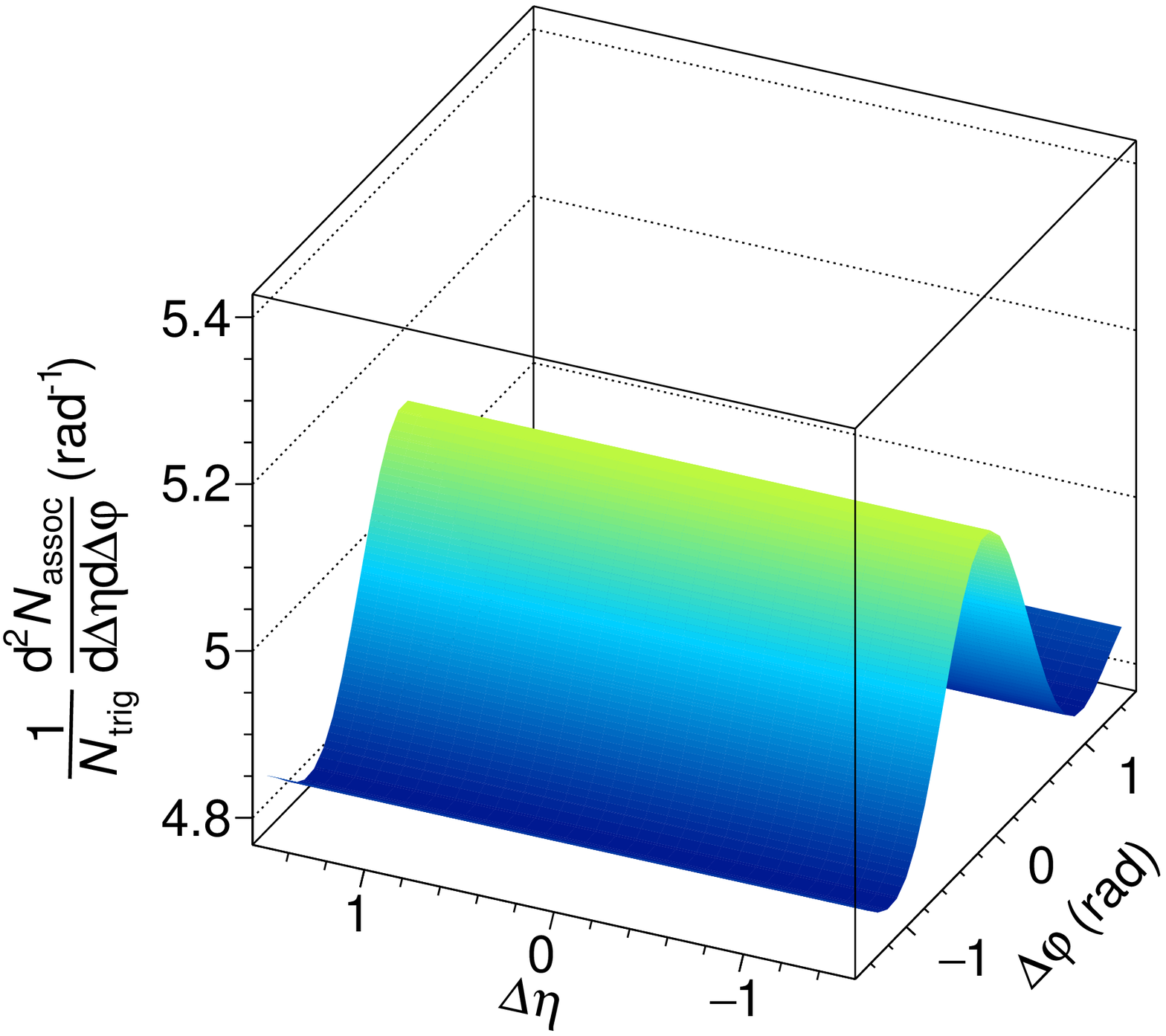} \label{subfig:fit2}}
\subfigure[][]{\includegraphics[width=0.45\textwidth]{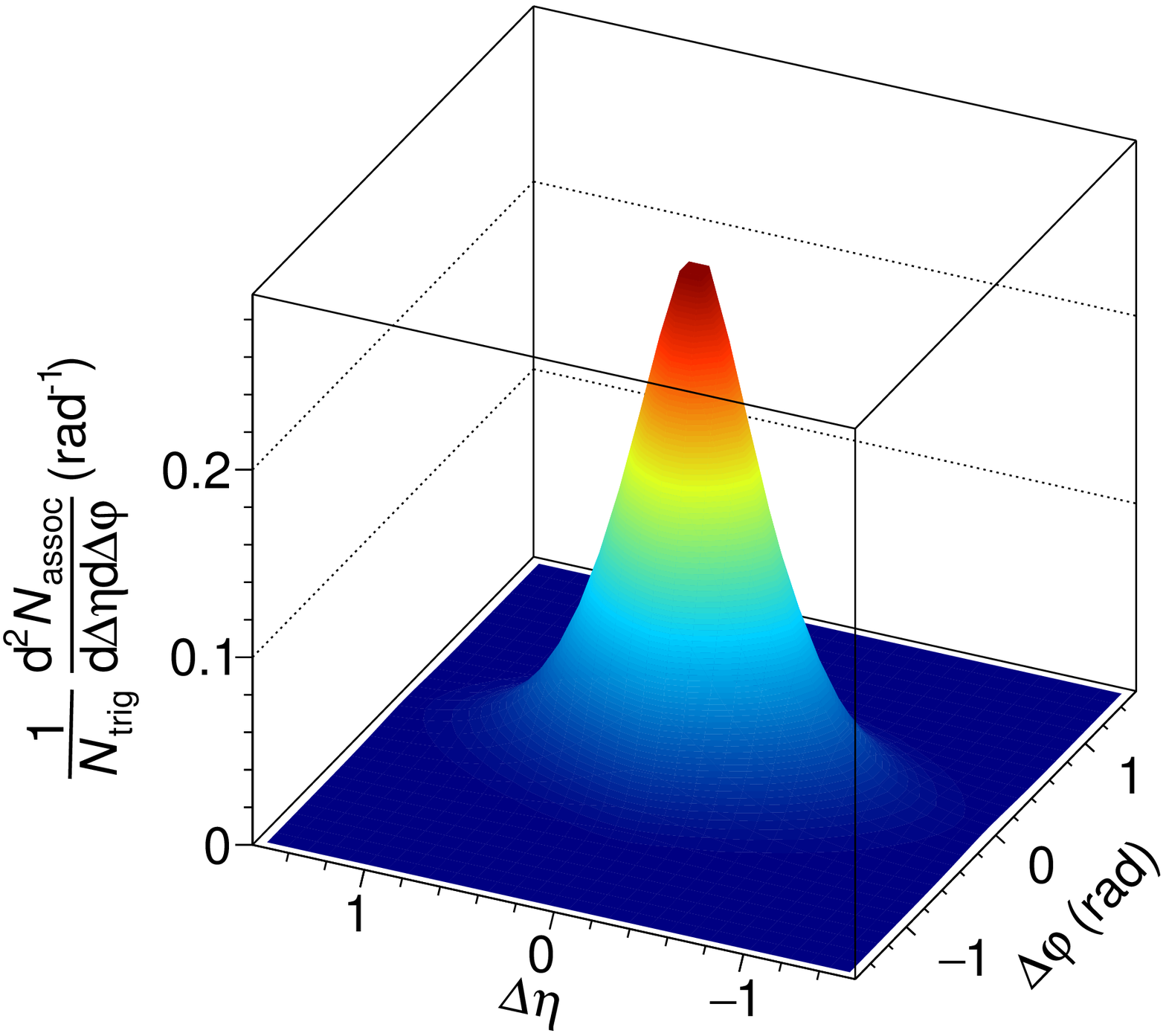} \label{subfig:fit3}}
\subfigure[][]{\includegraphics[width=0.45\textwidth]{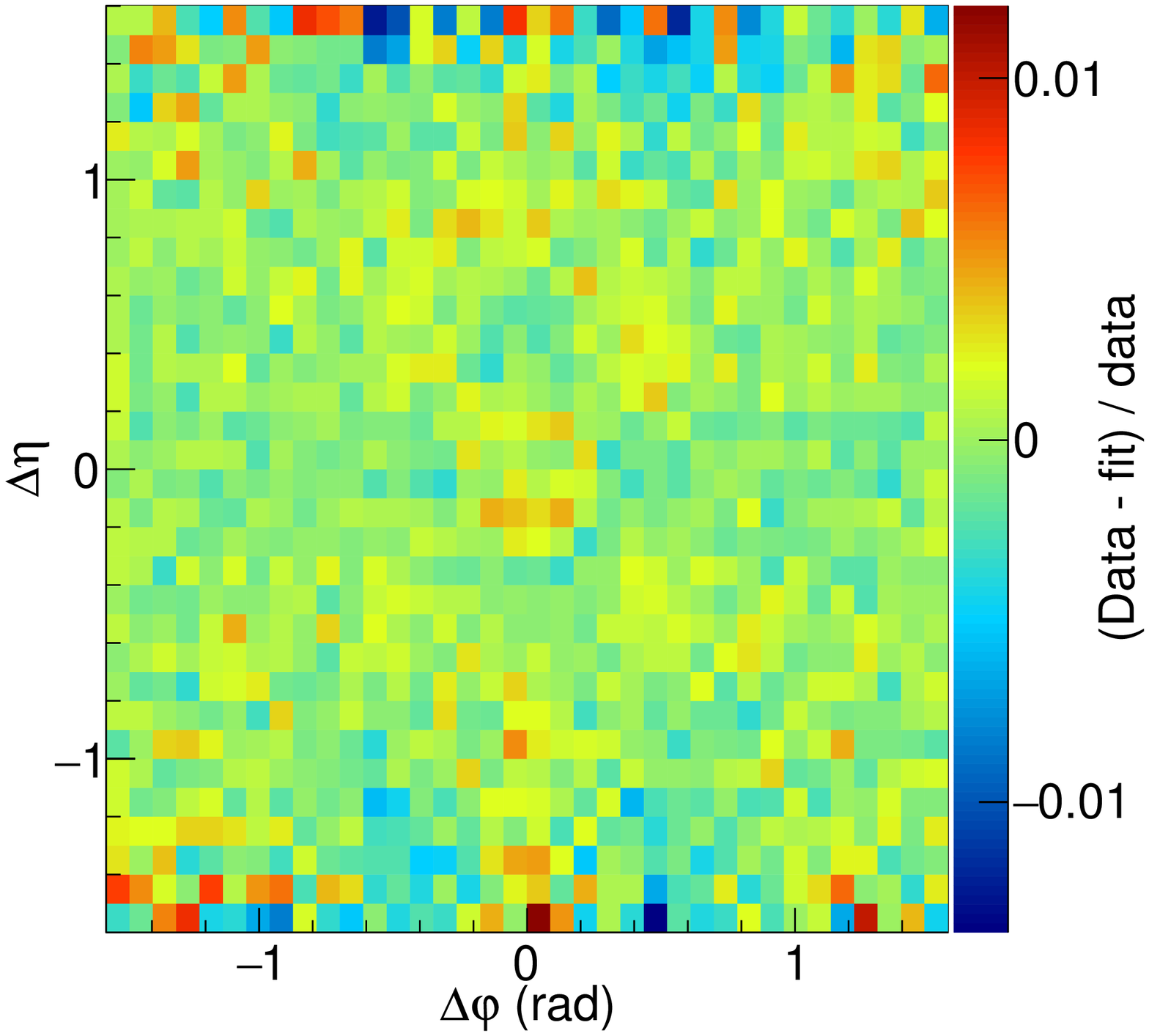} \label{subfig:fit4}}
\caption{\label{fig:fitexample} Illustration of the fitting procedure for the 10\% most central Pb--Pb events at \snn\ = \unit[2.76]{TeV} in $2<\pta<$~\unit[3]{\gevc} and $3<~\ptt<$~\unit[4]{\gevc}. The panels show \protect\subref{subfig:fit1} the two-dimensional azimuthal and pseudorapidity total per-trigger yield, \protect\subref{subfig:fit2} the background distribution  and \protect\subref{subfig:fit3} the signal peak component from the fit by Eq.~\ref{eq:fit}, and \protect\subref{subfig:fit4} the relative difference between the data and the fit.}
\end{figure*}

In order to characterize the near-side peak shape, a simultaneous fit of the peak, the combinatorial background and the long-range correlation background stemming from collective effects is performed. This exploits that in two-particle correlations the near-side peak is centred around $\Dphi = 0,~\Deta = 0$ while long-range correlation structures are approximately independent of $\Deta$~\cite{Adam:2016ows}. This strategy limits the analysis to the near side, as the away-side peak is elongated in $\Deta$. The fit function used is a combination of a constant, a generalized two-dimensional Gaussian function and $\cos( n \Dphi)$ terms for $n = 2, 3, 4$.
\begin{eqnarray}
  F(\Dphi,\Deta) &=& C_1 + \sum_{n=2}^4 2 V_{n\Delta} \cos (n \Dphi) + \nonumber \\
    && C_2 \cdot G_{\gamma_{\Dphi},w_{\Dphi}}(\Dphi) \cdot G_{\gamma_{\Deta},w_{\Deta}}(\Deta) \label{eq:fit} \\
    G_{\gamma_x,w_x}(x) &=& \frac{\gamma_x}{2{w}_x\Gamma (1/\gamma_x)} \exp \left[ -\left(\frac{|x|}{w_x}\right)^{\gamma_x} \right]
\end{eqnarray}
Thus, in Pb--Pb collisions, the background is characterized by 4 parameters ($C_1$, $V_{n\Delta}$) where $V_{n\Delta}$ are the Fourier components of the long-range correlations~\cite{Aamodt:2011by}, and it should be noted that the inclusion of  orders higher than 4 does not significantly change the fit results. In pp collisions, however, the background consists effectively only of the pedestal $C_1$. The peak magnitude is characterized by $C_2$, and the shape which is the focus of the present analysis by 4 parameters ($\gamma_{\Dphi},~w_{\Dphi},~\gamma_{\Deta},~w_{\Deta}$). Note that for $\gamma=2$ the generalized Gaussian function $G$ is a Gaussian, and for $\gamma=1$ it is a Laplace distribution, which is an exponential where the absolute value of the argument is taken ($\exp(-|x|)$). The aim of using this fit function is to allow for a compact description of the data rather than attempting to give a physical meaning to each parameter. A further reduced description of the peak shape is provided by the variances ($\sigma_{\Dphi}$ and $\sigma_{\Deta}$) of the generalized Gaussian. The evolution of the peak shape from peripheral to central collisions is described by the ratio of the width in the central bin (0--10\%) and the peripheral bin (\mbox{50--80\%}), denoted by $\sigma^{\rm CP}_{\Dphi}$ and $\sigma^{\rm CP}_{\Deta}$.

In the data, a depletion around $\Dphi = 0$, $\Deta = 0$ is observed at low $\pt$, however, the fit function does not include such a depletion. Several bins in the central region are excluded from the fit avoiding a bias on the extracted peak width. The size of the excluded region varies with $\pt$ and collision centrality  reflecting both the width of the peak and the area of the depletion. The exclusion region is largest (0.3) in the lowest $\pt$ bin and most central Pb--Pb collisions and vanishes for higher $\pt$ and peripheral Pb--Pb collisions. The sensitivity of the result to the size of the exclusion region was studied, see Section~\ref{sec:systunc}. Thus, by definition, the peak width describes the shape of the peak outside of the central region. The depletion in the central region is quantified by the near-side depletion yield in Section~\ref{sec:dip} by computing the difference between the fit and the per-trigger yield within the exclusion region.

\begin{figure*}[t!]
\centering
\subfigure[][]{\includegraphics[width=0.45\textwidth]{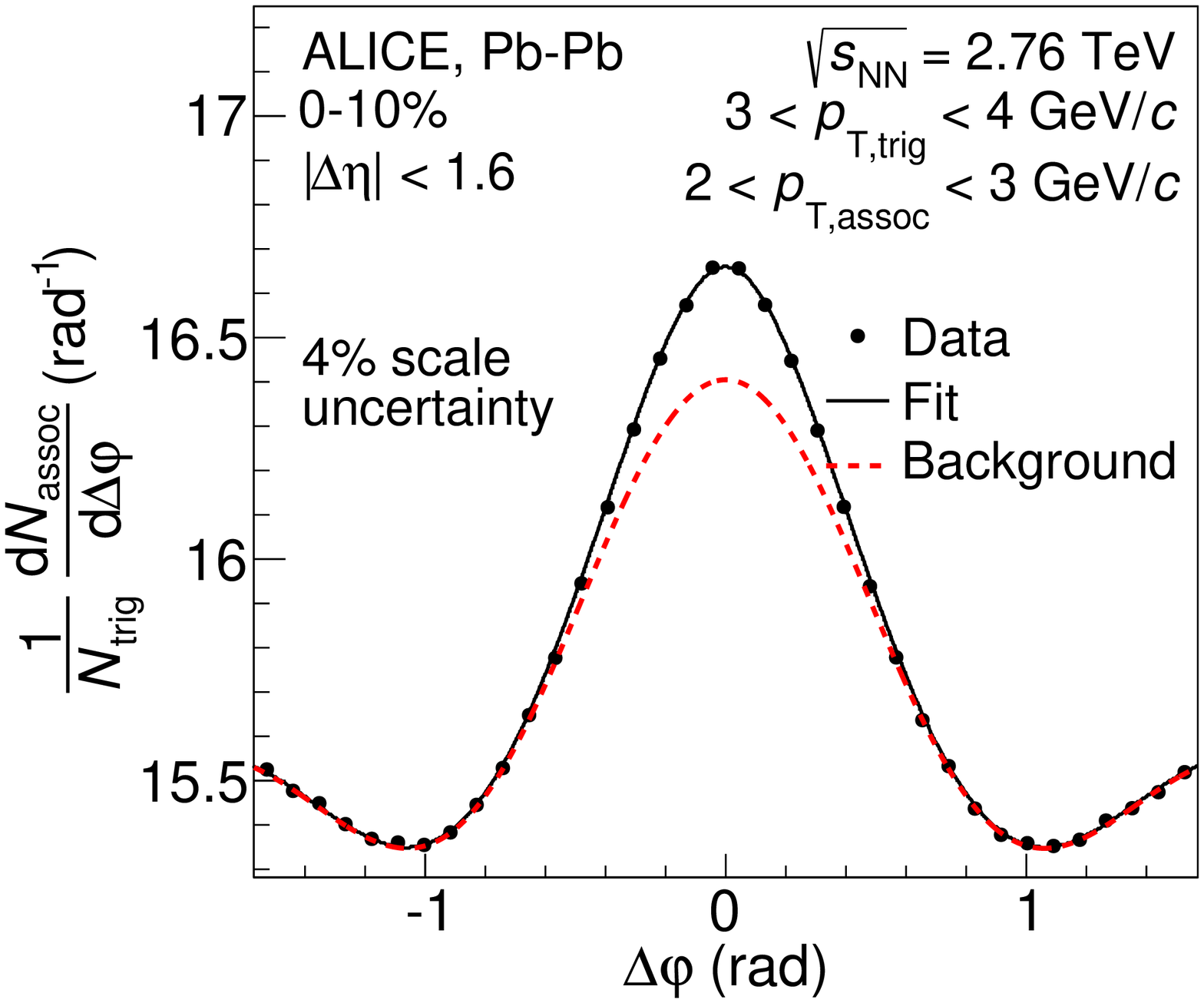} \label{subfig:results2c_projPhi}}
\subfigure[][]{\includegraphics[width=0.45\textwidth]{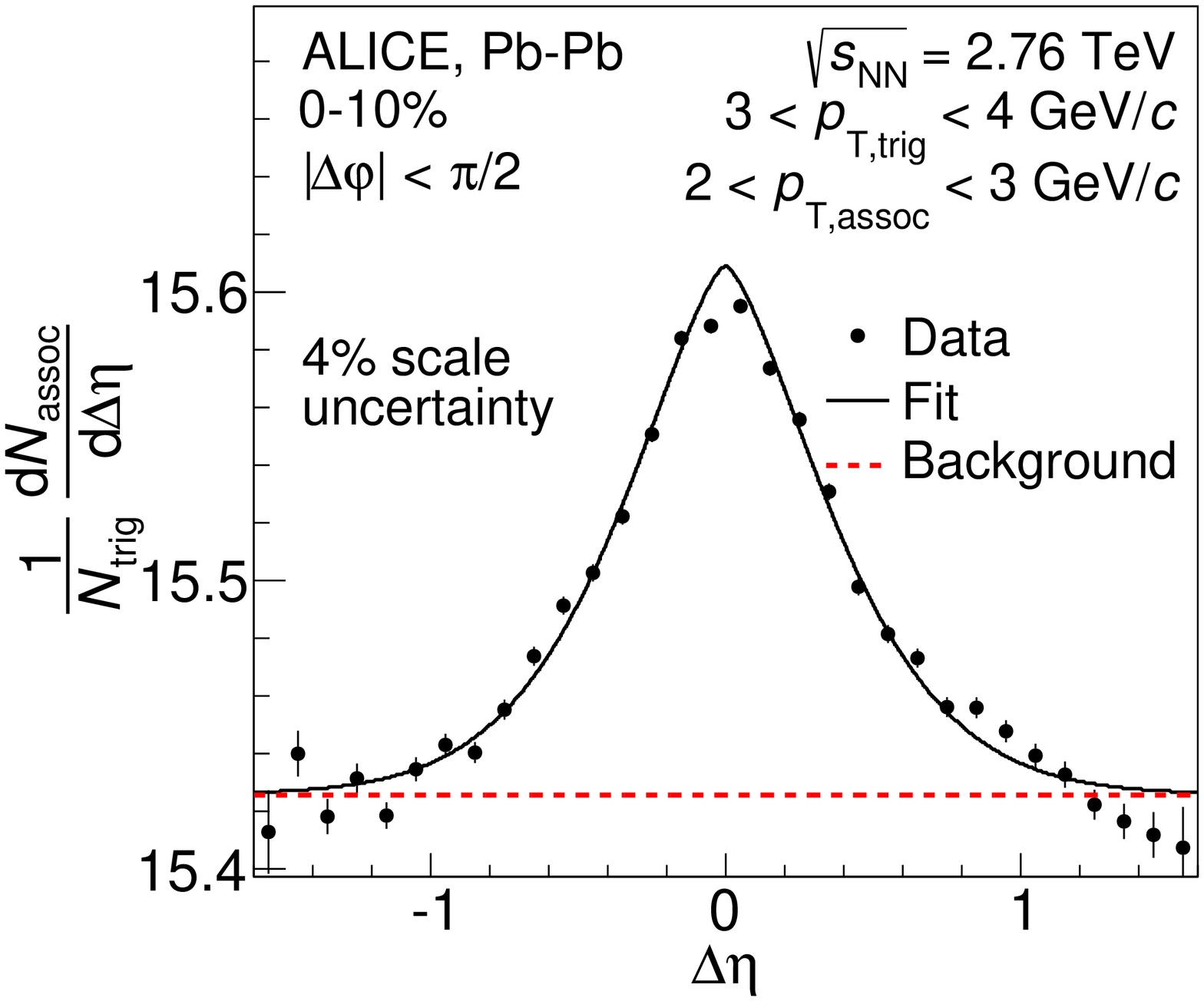} \label{subfig:results2c_projEta}}
\caption{\label{fig:proj}
Projections of Fig.~\ref{fig:fitexample} \subref{subfig:fit1} to the $\Dphi$ \protect\subref{subfig:results2c_projPhi} and $\Deta$ \protect\subref{subfig:results2c_projEta} directions. The projections integrated over $|\Deta| < 1.6$ and $|\Dphi| < \pi/2$, respectively, present per-trigger yields (and not densities) and therefore the level of the background is different than in Fig.~\ref{fig:fitexample}. The fit and the background component of the fit are overlaid with the data.}
\end{figure*}

Figure~\ref{fig:fitexample} illustrates the fit procedure. Shown is the data as well as the background and peak components of the fit. The bottom right panel shows the difference between the data and the fit where only minor deviations less than 0.5\% can be observed. Figure~\ref{fig:proj} shows the $\Dphi$ and $\Deta$ projections of the data overlaid with the obtained fit functions. The comparison with the background illustrates the magnitude of the peak.

In Pb--Pb collisions, the $\chi^2/{\rm ndf}$ values of the fits are found in the range 1.0--2.5; most are around 1.5. In the highest two $\pt$ bins (i.e. in $3<\pta<$~\unit[8]{\gevc} and $4<\ptt<$~\unit[8]{\gevc}) the values increase up to about 2.5 showing that at high $\pt$ the peak shape starts to depart from the generalized Gaussian description. In pp collisions, the $\chi^2/{\rm ndf}$ values are in the range 1.3--2.0.  

Different fitting strategies have been tried using a two-dimensional Gaussian to describe the peak, which is found to not describe the data satisfactorily (conversely, the $\chi^2/{\rm ndf}$ is too large). A superposition of two two-dimensional Gaussians describes the data well, but is found unstable compared to the generalized Gaussian. In general, the fit with a single two-dimensional Gaussian results in smaller peak widths than the generalized Gaussian case which in turn has smaller peak widths than the two two-dimensional Gaussian fit.

\section{Systematic uncertainties}
\label{sec:systunc}

Systematic uncertainties connected to the measurement are determined modifying the selection criteria discussed above and repeating the analysis.
 The difference in the extracted parameters is studied as a function of $\pt$, centrality and collision system, but these dependencies are rather weak and one uncertainty value can be quoted for each source of systematic uncertainty in most cases. Finally, the contribution from the different sources of systematic uncertainties are added in quadrature. The extracted peak widths are rather insensitive to changes in the selections (total uncertainty of about 2--4.5\%) while the near-side depletion yield defined in Sec.~\ref{sec:dip} is more sensitive (about 24--45\% uncertainty).

Table~\ref{tab:systematics} summarizes the different sources of systematic uncertainties which have been considered. Changes of vertex range and track selection have already been detailed in Sec.~\ref{sec:selection}. The selection criterion on pairs with small opening angles (see Sec.~\ref{sec:twopartfunc}) is increased by a factor 2 and the mass range in the cut removing neutral-particle decays is  modified by 50\%. The size of the exclusion region around $\Dphi = 0$, $\Deta = 0$ (see Sec.~\ref{sec:twopartfunc}) is enlarged by $0.17~(0.2)$ in the $\Dphi~(\Deta)$ direction. The sensitivity of the analysis results to the pseudorapidity range used is assessed by changing it by $\pm0.1$. This uncertainty includes effects of the pseudorapidity dependence of the anisotropic flow as well as the particle production in general. Trigger particles in positive and negative $\eta$-direction are studied separately to exclude any detector effects related to the trigger-particle direction. No dependence of the results presented in this paper on the polarity of the magnetic field was observed.

The influence of resonance decays on the observations presented below were investigated by performing the analysis separately for like-sign and unlike-sign pairs. While the numerical values change, which is not unexpected, the qualitative conclusions presented below are unchanged. In particular, the reported broadening and depletion are larger in the like-sign case suggesting that resonance decays do not play a significant role for these phenomena.

\begin{table*}[tb!] \centering
  \begin{tabular}{l|cc|cc|c} \hline
    Source				& $\sigma_{\Delta\varphi}$ & $\sigma_{\Delta\eta}$ & $\sigma^{CP}_{\Delta\varphi}$ & $\sigma^{CP}_{\Delta\eta}$ & Depletion yield \\ \hline
    Track selection and efficiencies    & \multicolumn{2}{c|}{1.0\%}     & \multicolumn{2}{c|}{1.3\%}     & 20\% \\
    Small opening angle cut		& \multicolumn{2}{c|}{0.7\%} 	 & \multicolumn{2}{c|}{1.3\%}     & 5--10\% \\
    Neutral-particle decay cut		& \multicolumn{2}{c|}{0.1\%} 	 & \multicolumn{2}{c|}{0.2\%}     & 8--20\% \\
    Vertex range                        & \multicolumn{2}{c|}{1.0\%}     & \multicolumn{2}{c|}{1.0\%}     & 5--10\% \\
    Pseudorapidity dependence           & 1.7\% & 4.1\%                 & 0.6\% & 2.5\%                 & 5--15\% \\
    Exclusion region			& 0.1\% & 1.0\%             	& 0.1\% & 1.5\%                 & 7--28\% \\ \hline
    Total				& 2.3\% & 4.5\%                 & 2.2\% & 3.6\%                 & 24--45\% \\\hline
  \end{tabular}
  \caption{\label{tab:systematics} Summary of the systematic uncertainties of the analysis. Ranges indicate a dependence on centrality.}
\end{table*}

\section{Results}
\label{sec:results}

\begin{figure*}[htpb!]
\centering
\subfigure[][]{\includegraphics[width=0.32\textwidth]{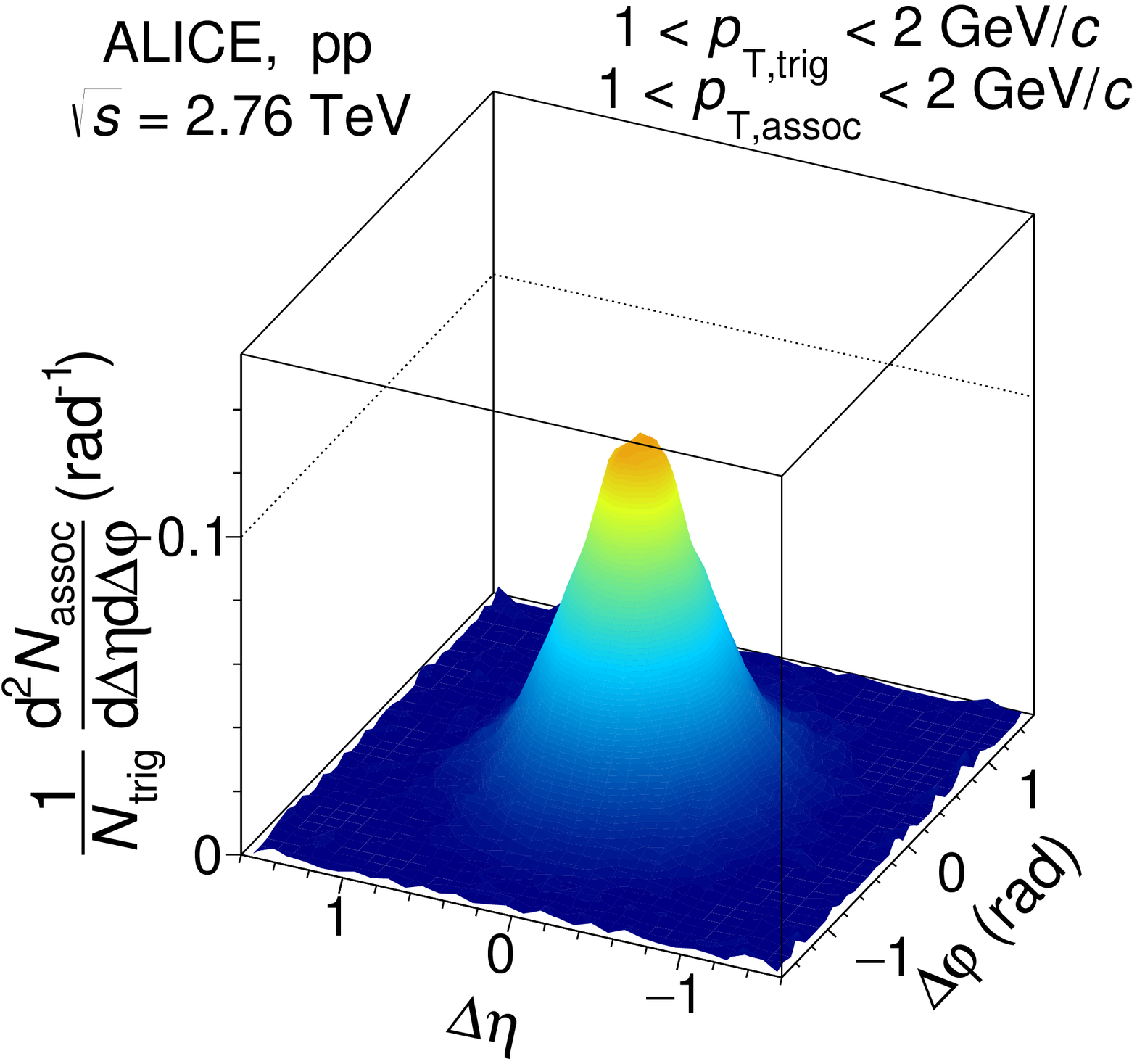}}
\subfigure[][]{\includegraphics[width=0.32\textwidth]{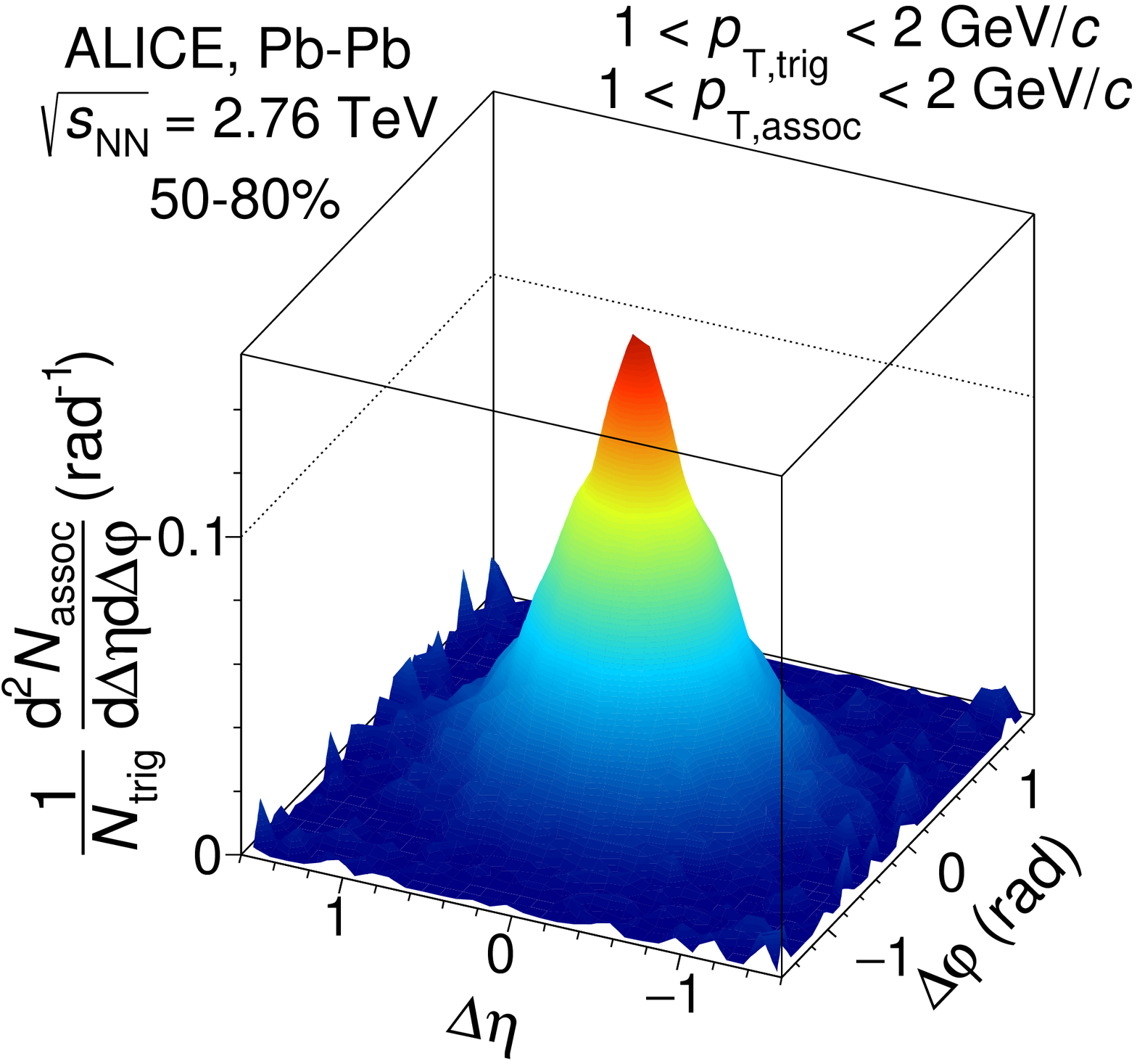}}
\subfigure[][]{\includegraphics[width=0.32\textwidth]{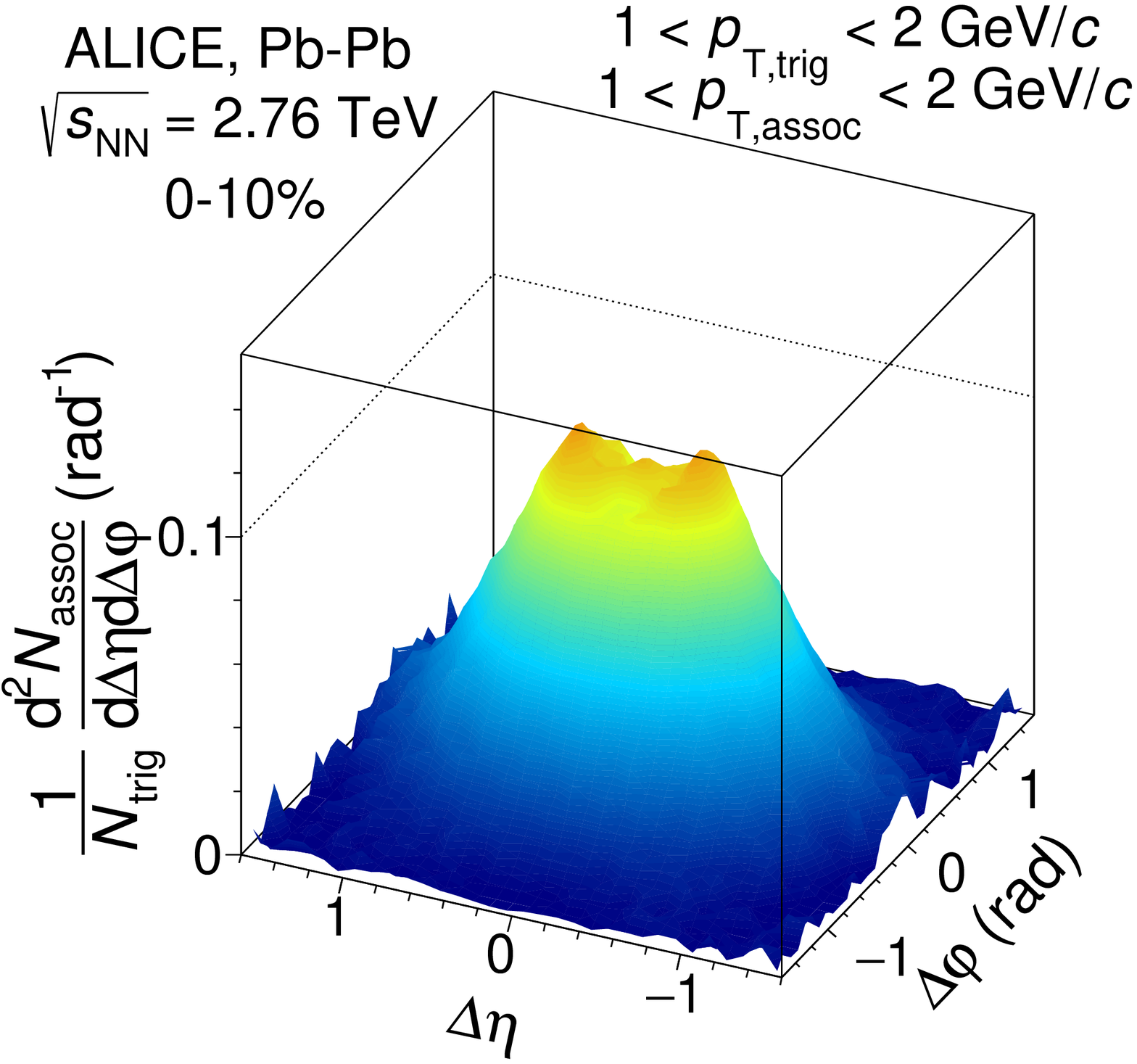} \label{subfig:results1c}}
\subfigure[][]{\includegraphics[width=0.32\textwidth]{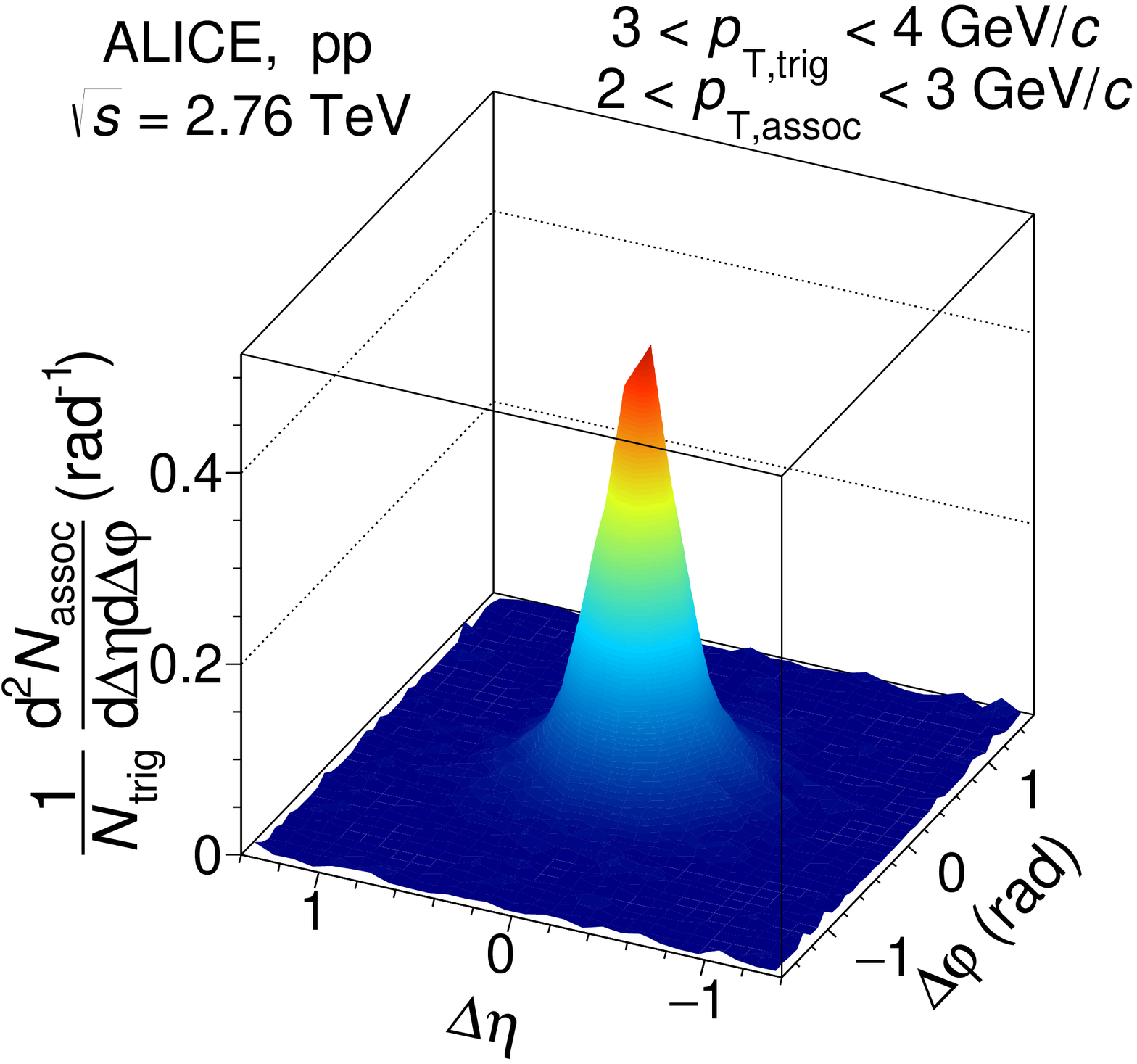}}
\subfigure[][]{\includegraphics[width=0.32\textwidth]{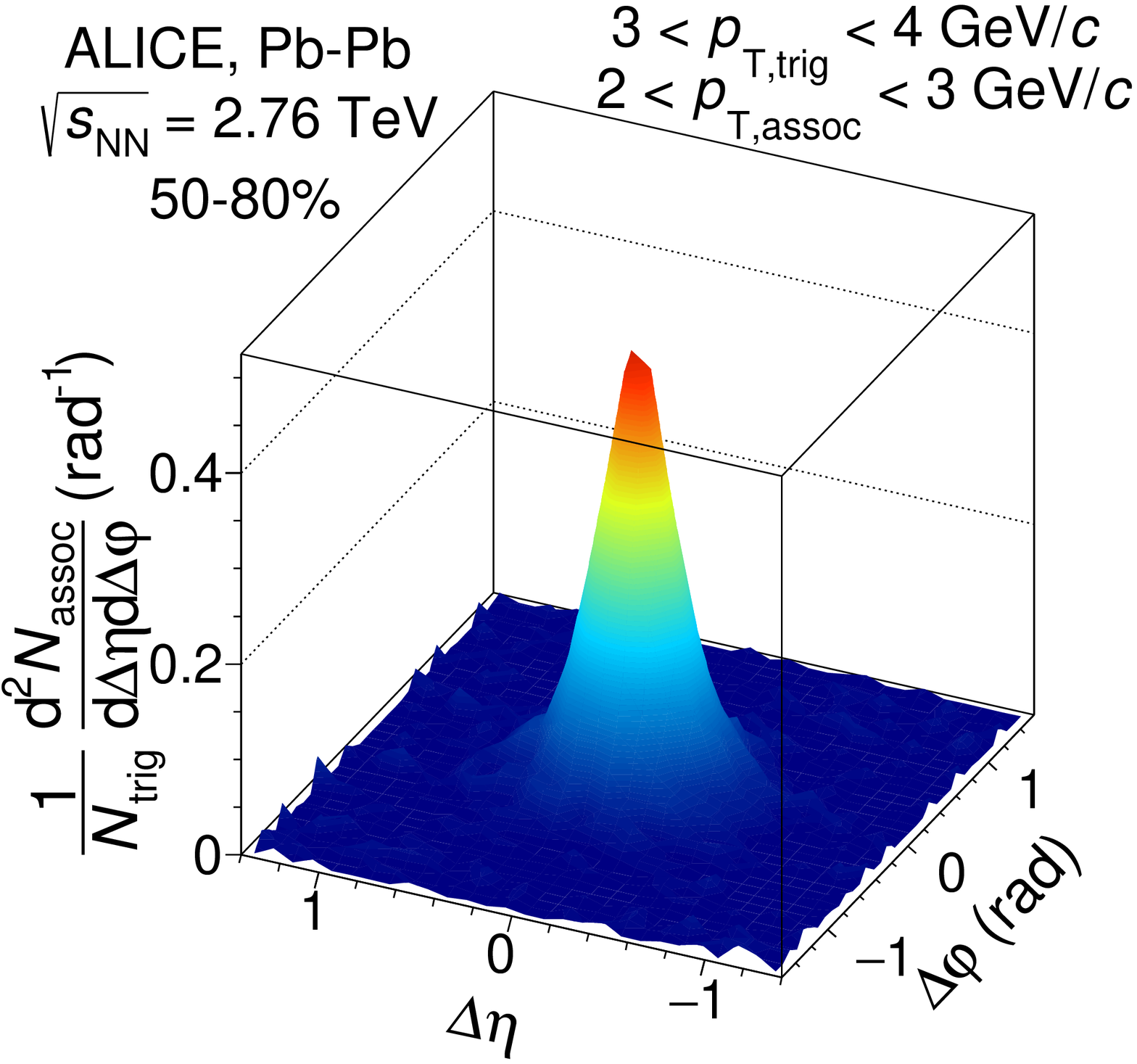}}
\subfigure[][]{\includegraphics[width=0.32\textwidth]{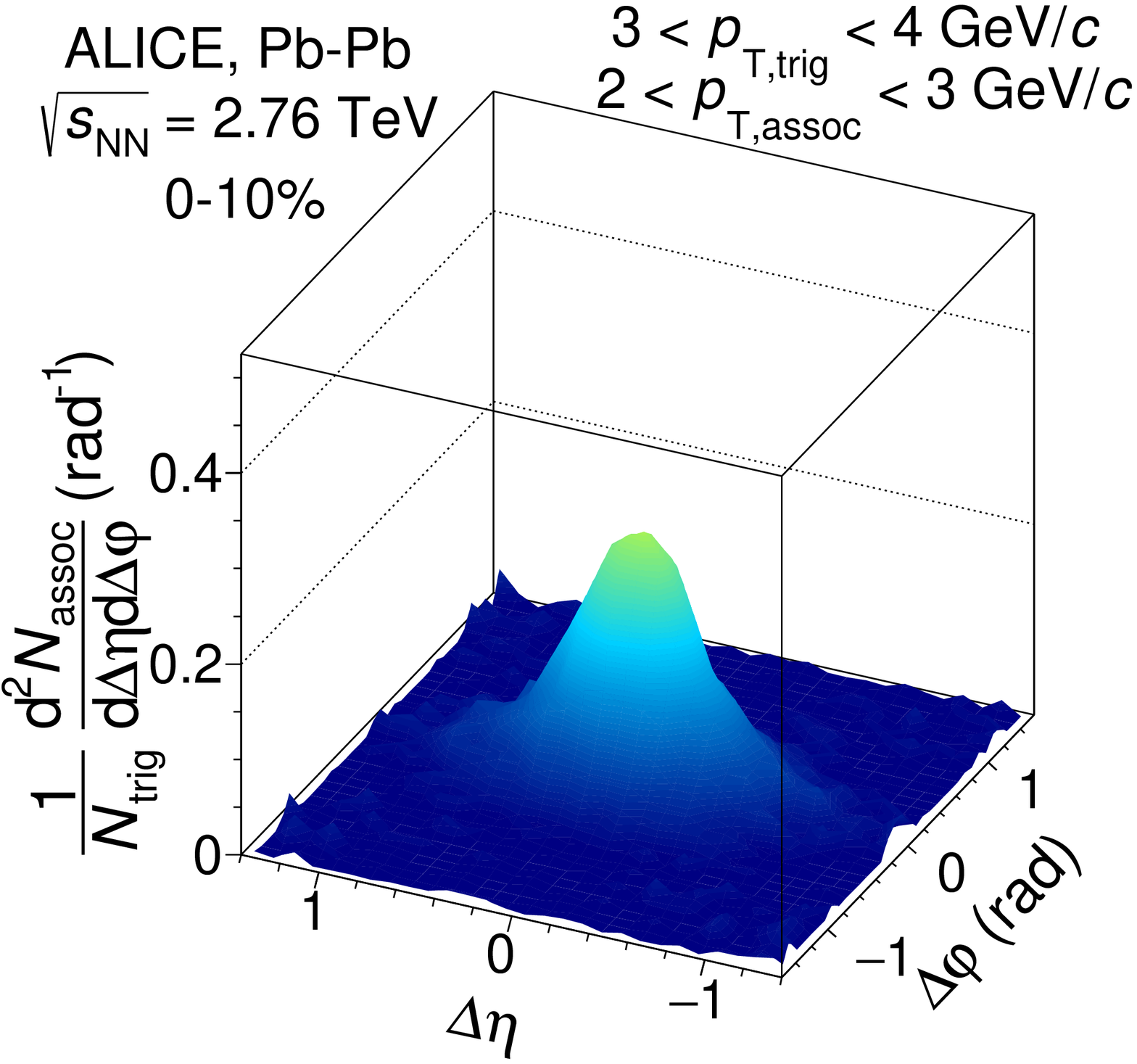}}
\caption{\label{fig:pertriggeryields}
Associated yield per trigger particle as a function of $\Dphi$ and $\Deta$ in pp collisions (left panels) and \mbox{Pb--Pb} collisions at \snn\ = \unit[2.76]{TeV} in the 50--80\% centrality class (middle panels) and in the 0--10\% centrality class (right panels). The top row shows $1<\pta<$~\unit[2]{\gevc} and $1<\ptt<$~\unit[2]{\gevc} and the bottom row shows $2<\pta<$~\unit[3]{\gevc} and $3<\ptt<$~\unit[4]{\gevc}. The background obtained from the fit function has been subtracted in order to emphasize the near-side peak.
}
\end{figure*}

\begin{figure*}[t!]
\centering
\subfigure[][]{\includegraphics[width=0.45\textwidth]{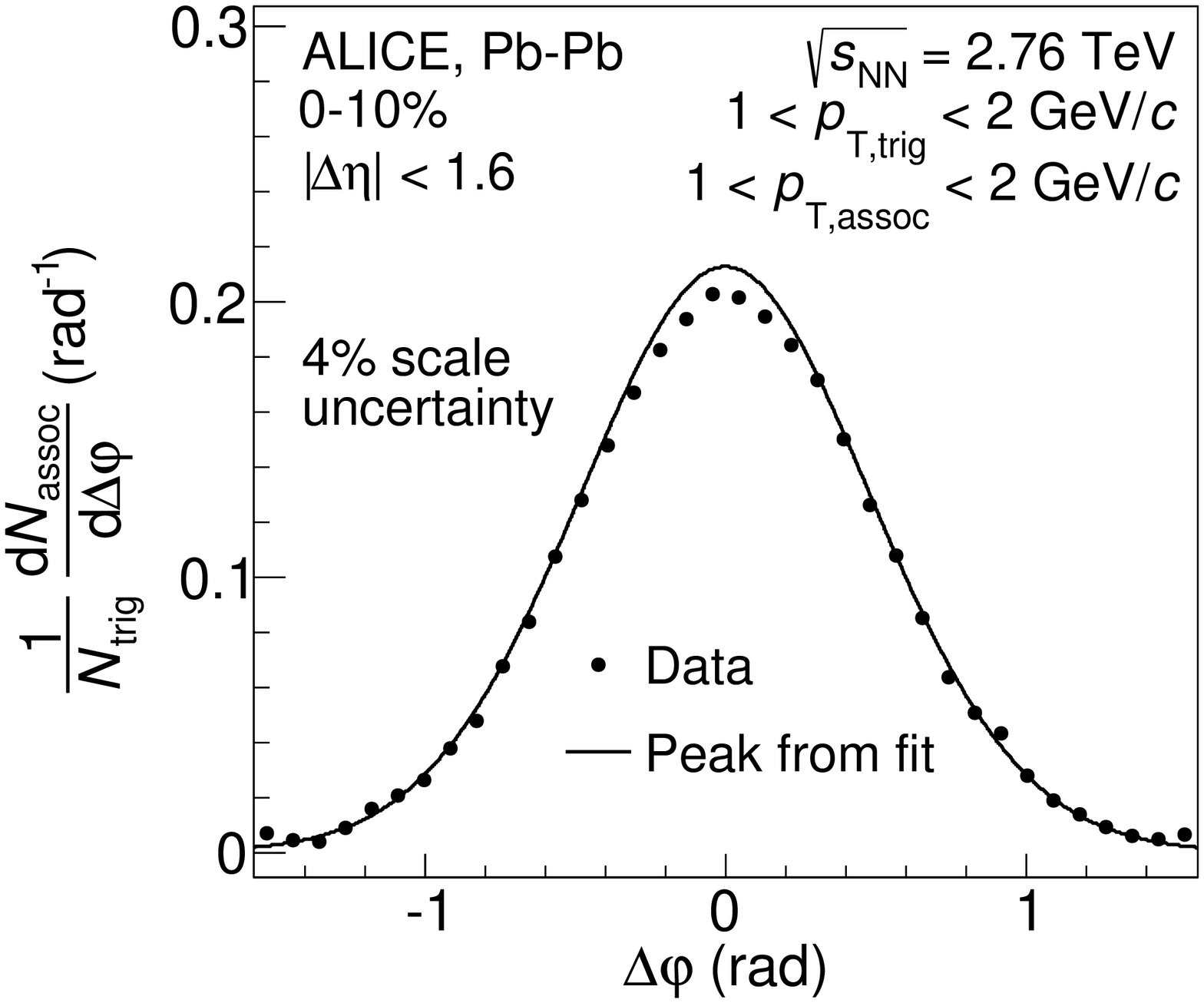} \label{subfig:BGSubtractedresults1c_projPhi}}
\subfigure[][]{\includegraphics[width=0.45\textwidth]{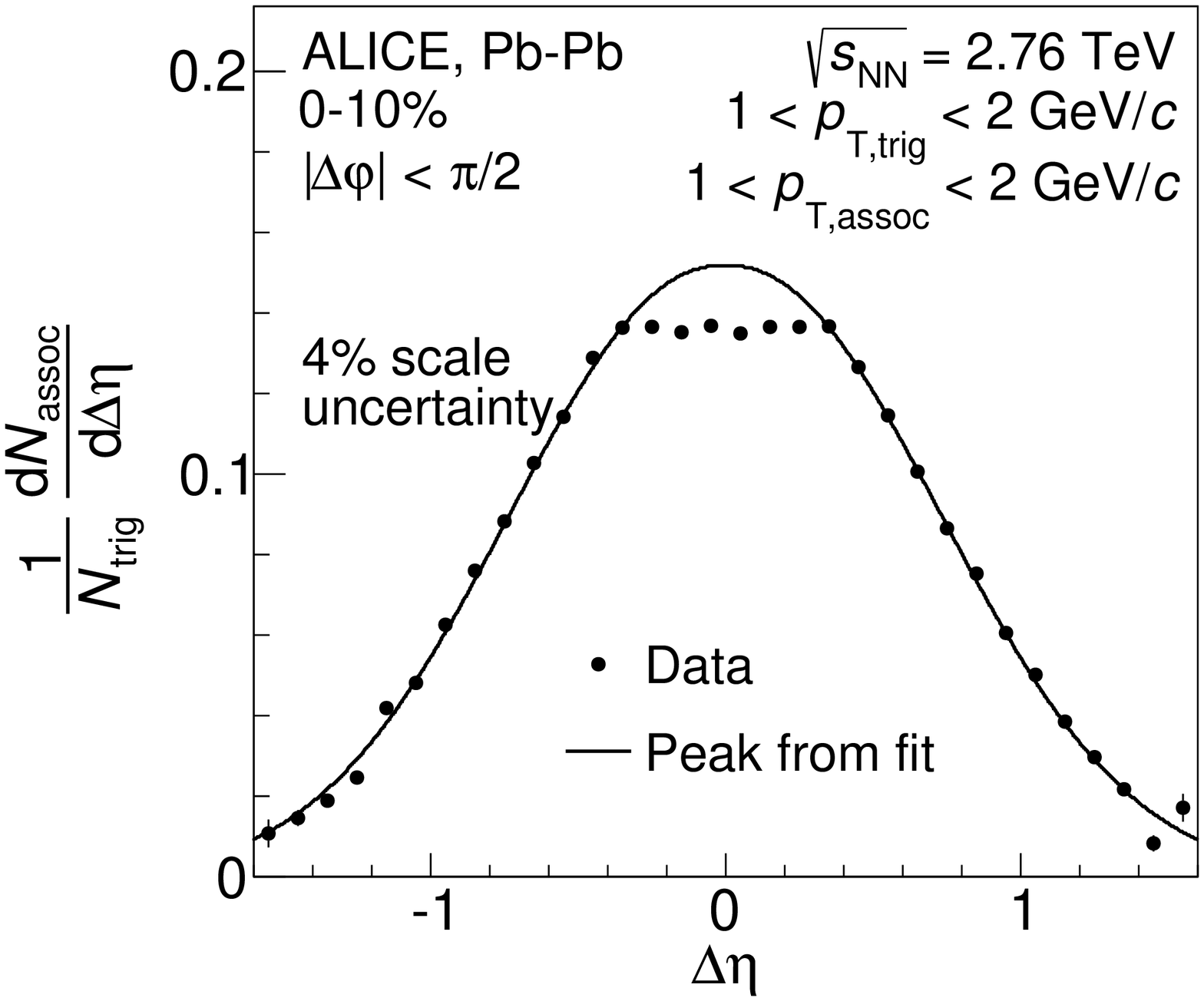} \label{subfig:BGSubtractedresults1c_projEta}}
\caption{\label{fig:projNoBG}
Projections of Fig.~\ref{subfig:results1c} to the $\Dphi$ \protect\subref{subfig:BGSubtractedresults1c_projPhi} and $\Deta$ \protect\subref{subfig:BGSubtractedresults1c_projEta} directions, the depletion around $\Dphi = 0$, $\Deta = 0$ is clearly visible in both directions.}
\end{figure*}

The top row of Fig.~\ref{fig:pertriggeryields} shows the near-side peak in $1<\ptt<$~\unit[2]{\gevc} and $1<\pta<$~\unit[2]{\gevc} after subtraction of the background estimated with Eq.~\ref{eq:fit}. The peak has a similar shape in pp collisions and in peripheral (50--80\% centrality) Pb--Pb collisions, where it is approximately symmetric in $\Dphi$ and $\Deta$. In the 10\% most central collisions a different picture is observed: the near-side peak is broader than in peripheral collisions and wider in $\Deta$ than in $\Dphi$. Furthermore, a depletion around $\Dphi = 0$, $\Deta = 0$ develops which will be discussed in more detail further below.
At higher $\pt$ (bottom row of Fig.~\ref{fig:pertriggeryields}), the near-side peak is also found broader in central collisions than in peripheral or pp collisions, although it is visually less pronounced, but the asymmetry between $\Dphi$ and $\Deta$ disappears at the  two highest $\pt$ bins included in the analysis. In addition, the amplitude of the peak is smaller in central collisions. Figure~\ref{fig:projNoBG} shows the projections of the two-dimensional histogram shown in Fig.~\ref{subfig:results1c}, where the depletion is largest, together with the fitted function.

\subsection{Peak widths}
\begin{figure*}[t!]
\centering
\includegraphics[width=\textwidth,trim=0 0 30 0,clip=true]{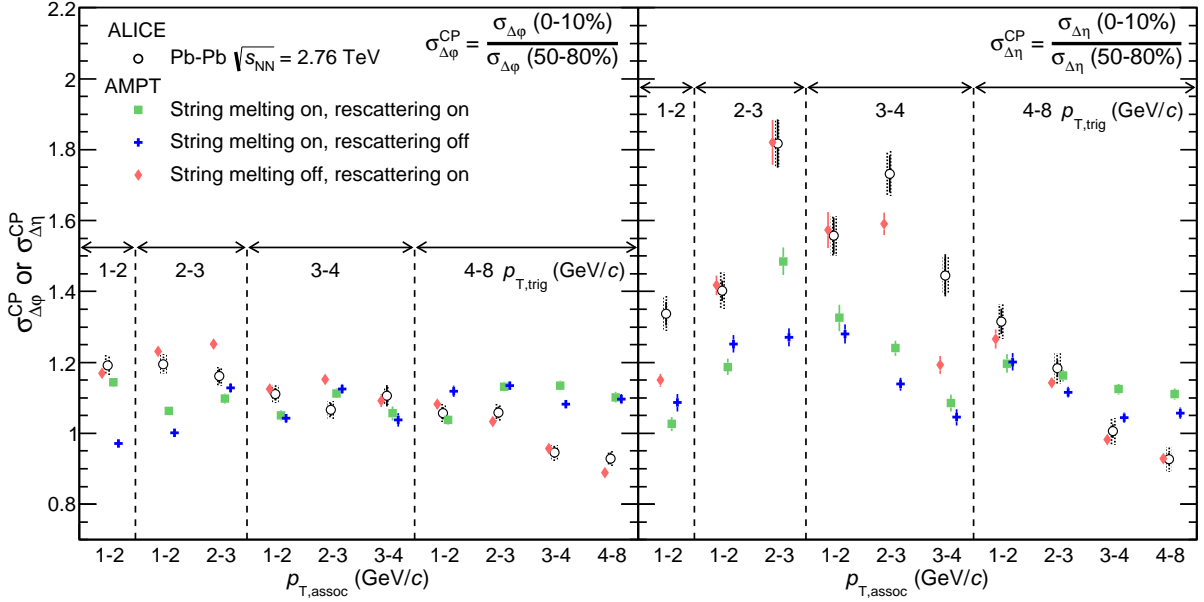}
\caption{\label{fig:widthratios}
Ratio of the peak widths in $\Dphi$ (left panel) and $\Deta$ (right panel) observed in central (0--10\%) and peripheral (50--80\%) collisions as a function of $\pt _{\rm ,trig}$ and $\pt _{\rm ,assoc}$ ranges. The data is compared to the different AMPT settings. Note that the $x$-axis combines the $\pta$ and $\ptt$ axis, and therefore, a uniform trend of the values is not expected. Lines indicate statistical uncertainties (mostly smaller than the marker size), while boxes (only for data) denote systematic uncertainties.
}
\end{figure*}

We examine and quantify the evolution of the near-side peak shape and width with the fit procedure described in Sec.~\ref{sec:twopartfunc}.
\begin{figure*}[t!]
\centering
\includegraphics[width=\textwidth,trim=0 0 30 0,clip=true]{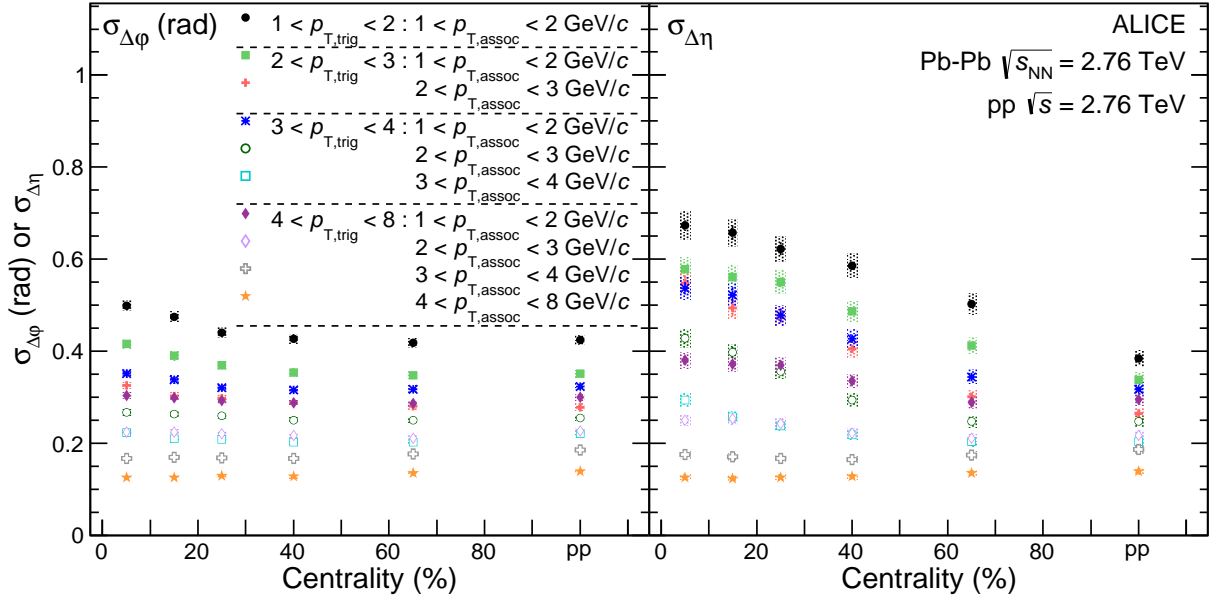}
\caption{\label{fig:sigma}
Shape parameters $\sigma_{\Dphi}$ (left panel) and $\sigma_{\Deta}$ (right panel) as a function of centrality in different $\pt$ ranges for Pb--Pb collisions at \snn\ = \unit[2.76]{TeV} and pp collisions (right most points in each panel). Lines indicate statistical uncertainties (mostly smaller than the marker size), while boxes denote systematic uncertainties. The markers are placed at the centre of the centrality bins.
}

\end{figure*}
The extracted shape parameters $\sigma_{\Dphi}$ and $\sigma_{\Deta}$ are presented in Fig.~\ref{fig:sigma}. In pp collisions, the $\sigma$ values range from 0.14 to 0.43 showing the expected $\pt$ dependence: due to the boost of the evolving parton shower at larger $\pt$ the peak is narrower.
In the $\Dphi$ direction (left panel) the values obtained in pp collisions are consistent with those in peripheral Pb--Pb collisions. The peak width increases towards central events which is most pronounced in the lowest $\pt$ bin (20\% increase). In the higher $\pt$ bins no significant width increase can be observed.
In the $\Deta$ direction (right panel) a much larger broadening towards central collisions is found. Already in peripheral collisions the width is larger than in pp collisions, and from peripheral to central collisions the width increases further up to $\sigma_{\Deta} = 0.67$ in the lowest $\pt$ bin. The largest relative increase of about 85\% is observed for $2<\ptt<$~\unit[3]{\gevc} and $2<\pta<$~\unit[3]{\gevc}. A significant broadening can be observed for all but the two largest $\pt$ bins. This increase is quantified for all $\pt$ bins in Fig.~\ref{fig:widthratios} by $\sigma^{\rm CP}_{\Dphi}$ and $\sigma^{\rm CP}_{\Deta}$. The increase is quantified with respect to peripheral Pb--Pb instead of pp to facilitate the MC comparisons discussed below.

In pp collisions, the peak shows circular symmetry in the $\Deta$--$\Dphi$ plane for all $\pt$. In Pb--Pb collisions, the peak becomes asymmetric towards central collisions for all but the two highest $\pt$ bins. The magnitude of this asymmetry depends on $\pt$ and is largest with about 70\% ($\sigma_{\Deta} > \sigma_{\Dphi}$) in the range $2<\ptt<$~\unit[3]{\gevc} and $2<\pta<$~\unit[3]{\gevc}.

\subsection{Model comparison}

The interplay of longitudinal flow with a fragmenting high $\pt$ parton 
was suggested in Ref.~\cite{Armesto:2004pt} as a possible source for
the observed asymmetric peak shape.
The authors argue that hard partons are interacting with a medium which shows collective behaviour. This is confronted with the simpler picture where the parton propagates through an isotropic medium with respect to the parton direction. In their calculation the scattering centres are Lorentz-boosted by applying a momentum shift depending on the collective component transverse to the parton-propagation direction.
The calculation in Ref.~\cite{Armesto:2004pt} for Au--Au collisions at \snn\ = \unit[200]{GeV} expects a 20\% increase from peripheral to central events for the $\Dphi$ direction and a 60\% increase for the $\Deta$ direction.
Despite the different centre-of-mass energy and collision system, the calculation is in quantitative agreement with the results presented in this paper.

Further studies on the possibility that the effect can be caused by an interplay of flow and jets have been done comparing the data
to generator-level results from A Multi-Phase Transport model (AMPT)~\cite{Lin:2004en, Xu:2011fi} which has been shown to feature a longitudinal broadening of the near-side peak \cite{Ma:2008nd}. Two mechanisms in AMPT produce collective effects: partonic and hadronic rescattering. 
Before partonic rescattering, the initially produced strings may be broken into smaller pieces by the so-called string melting.
Three different AMPT settings are considered,  having either string melting or hadronic rescattering or both activated.\footnote{AMPT versions v1.25t3 (without string melting, parameter isoft = 1) and v2.25t3 (with string melting, parameter \mbox{isoft = 4)} are used. In addition, in one sample the use of rescattering in the hadronic phase is disabled by setting the parameter ntmax to 3 (the default is 150). See Ref.~\cite{Adam:2016nfo} for more details on these settings.} About 10 million events were generated for each of the cases  with string melting  activated, and about 47 million events for the case  with string melting  disabled. The results obtained in pp collisions are compared to PYTHIA~8.1 simulations~\cite{Sjostrand:2007gs} with the Monash tune~\cite{Skands:2014pea} with about 500 million generated events.

The peak widths and $\sigma^{\rm CP}_{\Dphi}$ and $\sigma^{\rm CP}_{\Deta}$ are extracted from particle level AMPT simulations in the same way as for the data. Figure~\ref{fig:widthratios} compares these ratios to the data.
In the $\Dphi$ direction, the setting with string melting deactivated and hadronic rescattering active follows the trend of the data closest. The two other settings show a more uniform distribution across $\pt$ and only differ in the two lowest $\pt$ bins.
In the $\Deta$ direction, the setting with string melting deactivated and hadronic rescattering active quite remarkably follows the trend of the data including the large increase for intermediate $\pt$. The two other settings show qualitatively a similar trend but miss the data quantitatively.

\begin{figure*}[t!]
\centering
\subfigure[][]{\includegraphics[width=0.49\textwidth]{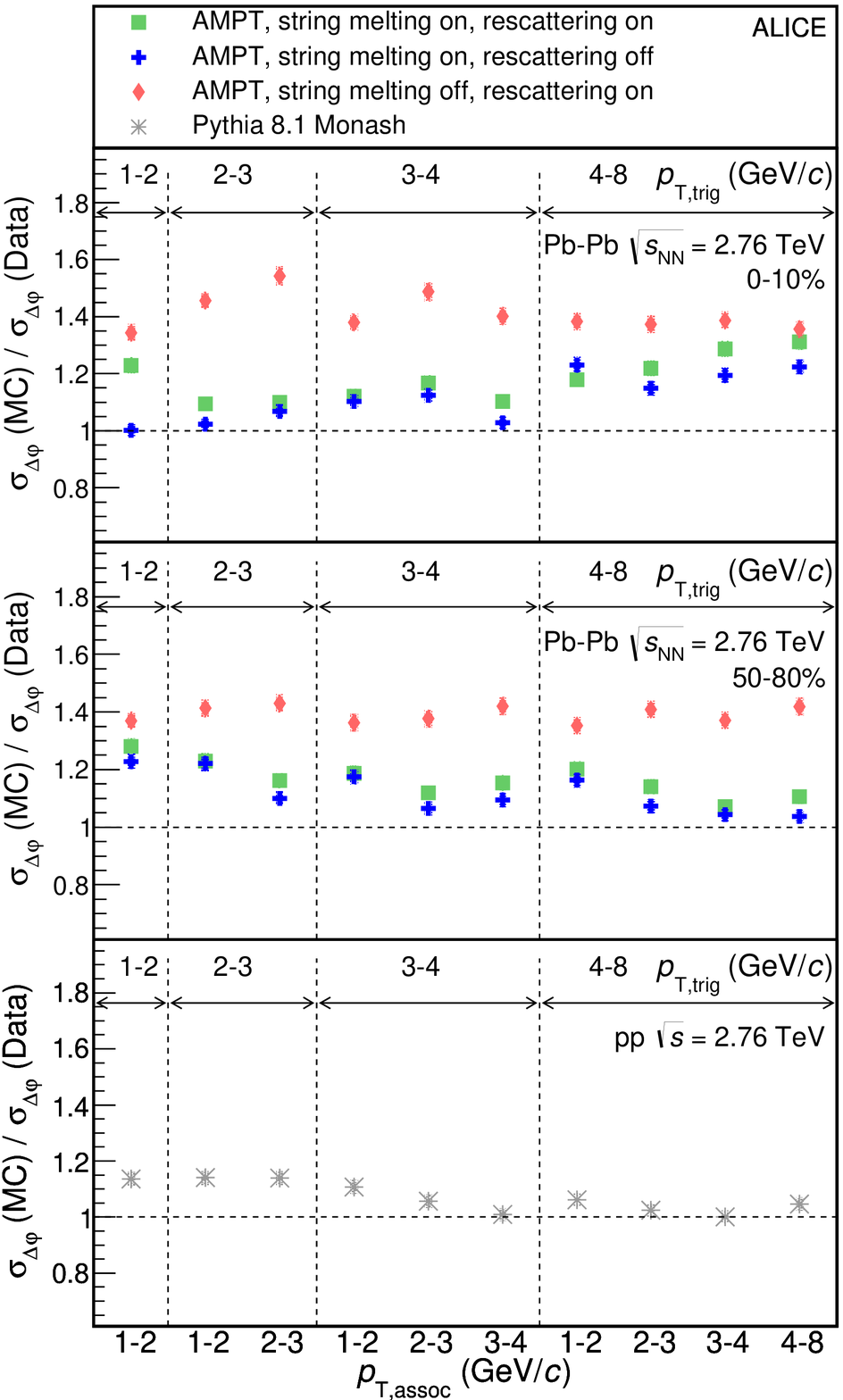} \label{subfig:absolute_ratio1a}}
\hfill
\subfigure[][]{\includegraphics[width=0.49\textwidth]{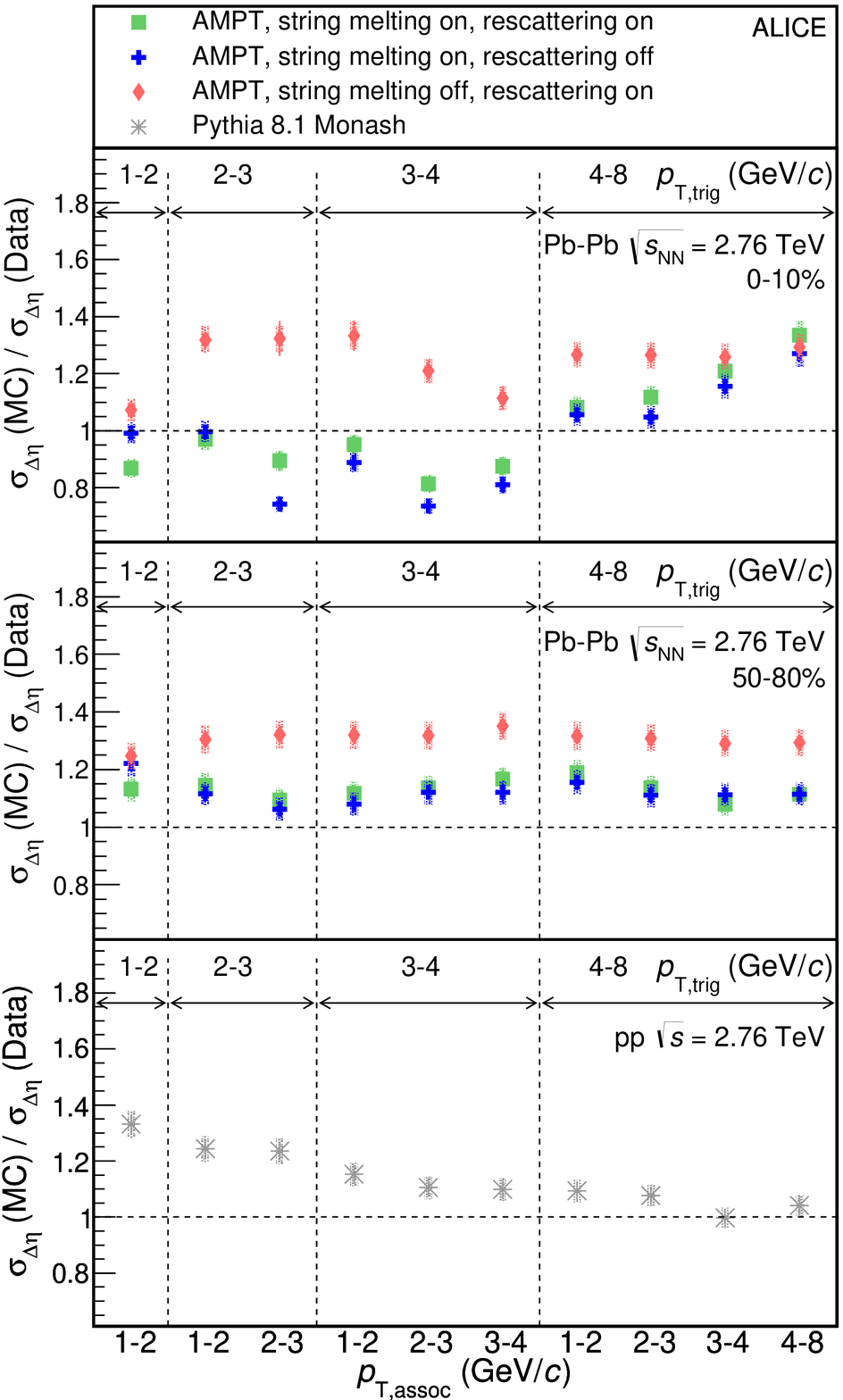} \label{subfig:absolute_ratio1b}}
\caption{\label{fig:ampt_ratio}
Ratio of the peak widths in AMPT to Pb--Pb collisions in central (top panels) and peripheral collisions (centre panels) as well as in  PYTHIA to pp collisions (bottom panels). Panel \protect\subref{subfig:absolute_ratio1a} show the shape parameters $\sigma_{\Dphi}$ while panel \protect\subref{subfig:absolute_ratio1b} show $\sigma_{\Deta}$. Lines indicate statistical uncertainties (mostly smaller than the marker size), while boxes denote systematic uncertainties.
}
\end{figure*}

In addition to the relative increase, it is interesting to compare the absolute widths. Figure~\ref{fig:ampt_ratio} presents the ratio of the widths in the three AMPT settings to the width measured in Pb--Pb collisions as well as the ones from PYTHIA simulations with the Monash tune to the ones measured in pp collisions.
In general, none of the AMPT settings provides an accurate description of the data. The setting which matches best the relative width increase (string melting deactivated, hadronic rescattering active), overestimates the width by on average 20--30\% with a mild $\pt$ dependence.
The two settings with string melting show a decreasing (increasing) trend as a function of $\pt$ in central (peripheral) collisions in the $\Dphi$ direction. In the $\Deta$ direction, in central collisions, they both over- and underestimate the data depending on $\pt$, while there is about 10\% overestimation in peripheral collisions mostly independent of $\pt$.
The width in pp collisions is well described by PYTHIA at high $\pt$ in both directions, while the width in $\Dphi$ ($\Deta$) is overestimated by 10\% (25\%) at low $\pt$.

\subsection{Near-side depletion}
\label{sec:dip}

The results presented in the previous section have focused on the overall shape of the near-side peak. In addition to the broadening, a distinct feature in central collisions and at the low $\pt$ is observed, a depletion around $\Dphi = 0$, $\Deta = 0$ (top right panel of Fig.~\ref{fig:pertriggeryields} and Fig.~\ref{fig:projNoBG}).

An extensive set of studies was carried out to determine whether this depletion could arise from detector effects. Studies focused, in particular, on two-track effects: tracks with similar momenta which overlap in parts of the detector volume may suffer from efficiency losses and reconstruction imperfections, e.g. a splitting of a particle's trajectory into two tracks may cause distortions of the two-particle correlation around $\Dphi = 0$, $\Deta = 0$. It was shown that such detector-related effects are present but only in a very limited region of where both $|\Dphi|$ and $|\Deta|$ are smaller than 0.04--0.05. The depletion discussed in this section extends out to $|\Deta|$ well beyond 0.3 which is significantly larger than the detector resolution and the reach of two-track efficiency effects. A detector-related origin is thus excluded.

\begin{figure*}[t!]
\centering
\subfigure[][]{\includegraphics[width=0.32\textwidth]{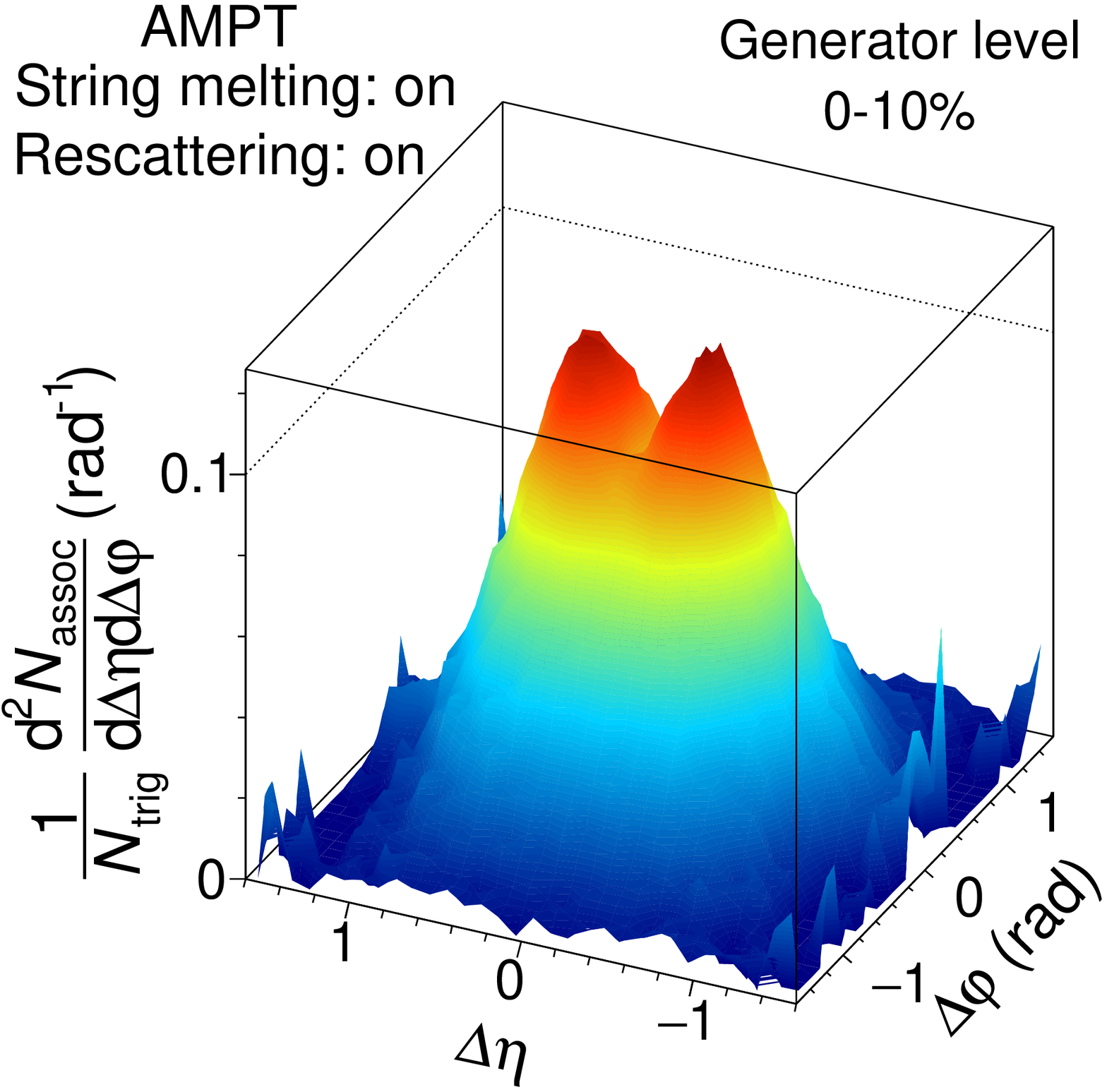}}
\subfigure[][]{\includegraphics[width=0.32\textwidth]{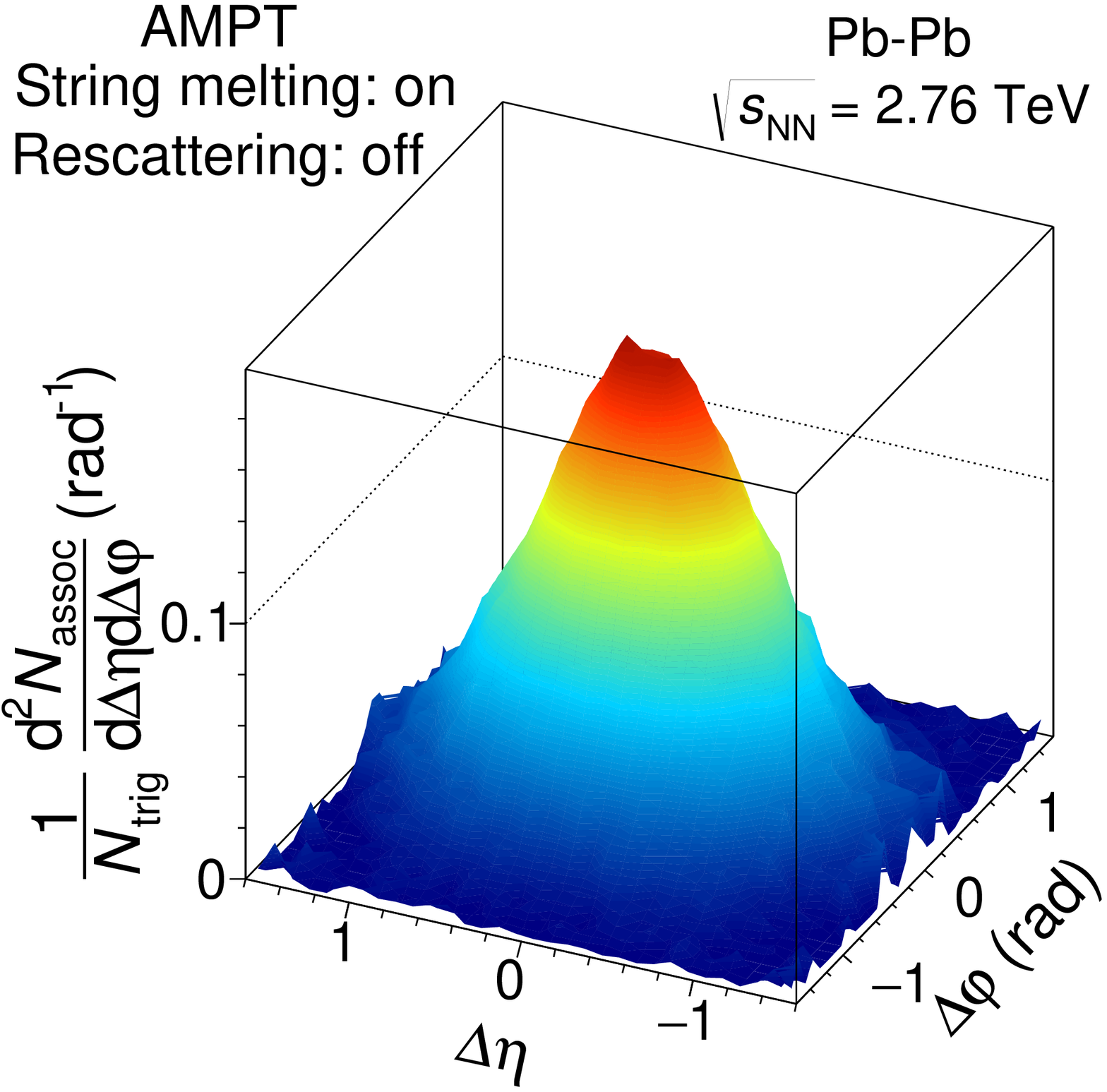}}
\subfigure[][]{\includegraphics[width=0.32\textwidth]{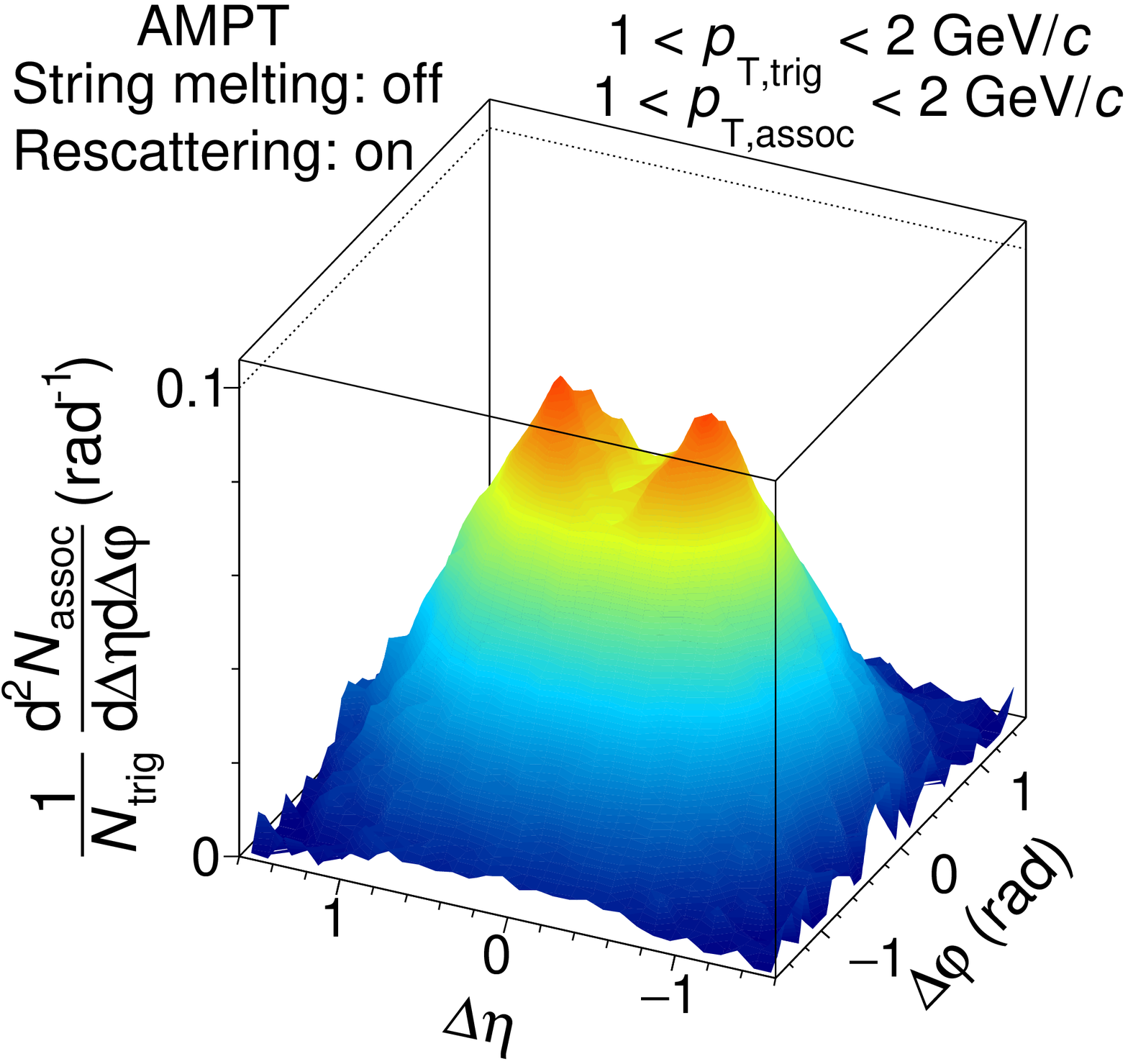}}
\subfigure[][]{\includegraphics[width=0.32\textwidth]{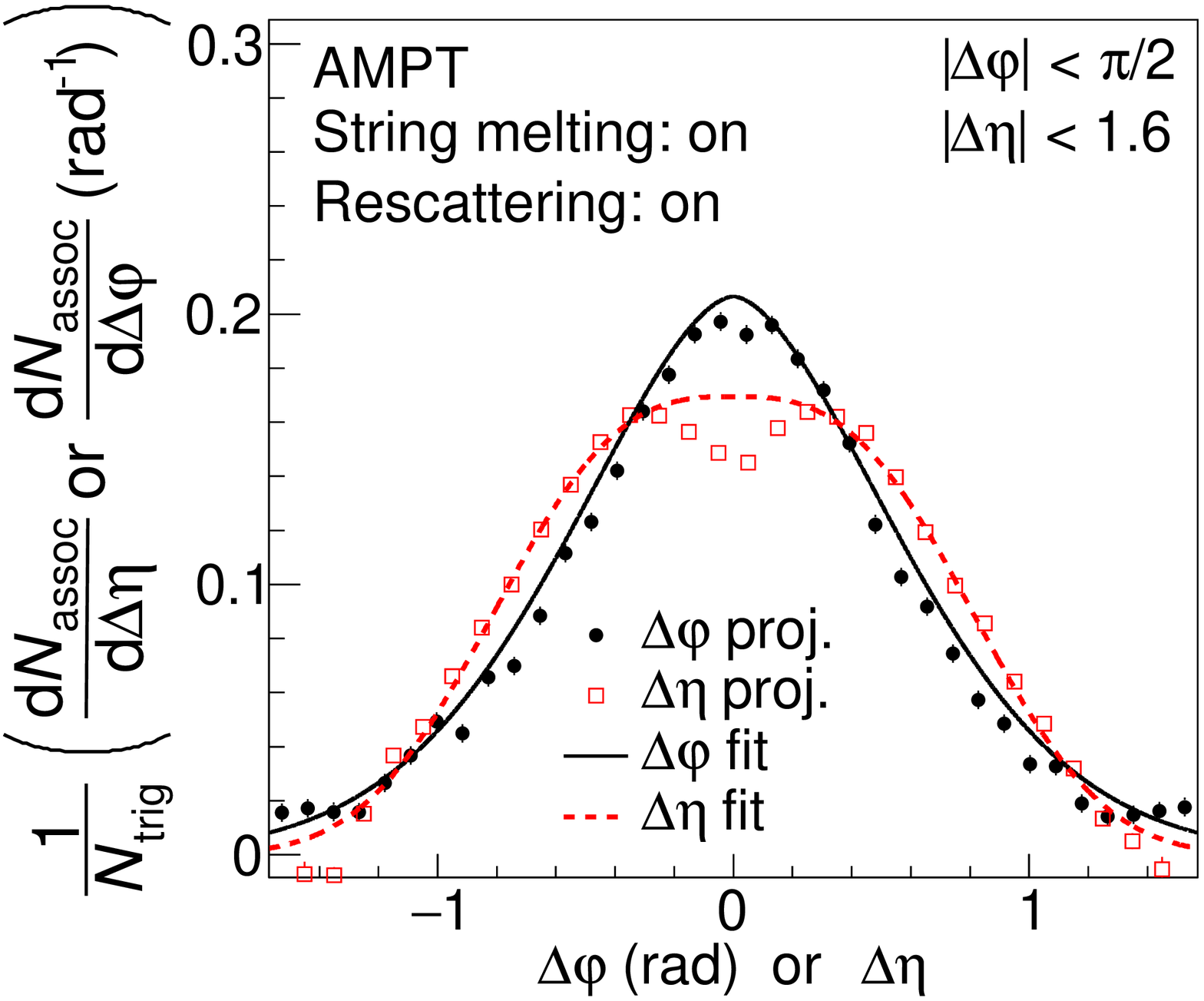}}
\subfigure[][]{\includegraphics[width=0.32\textwidth]{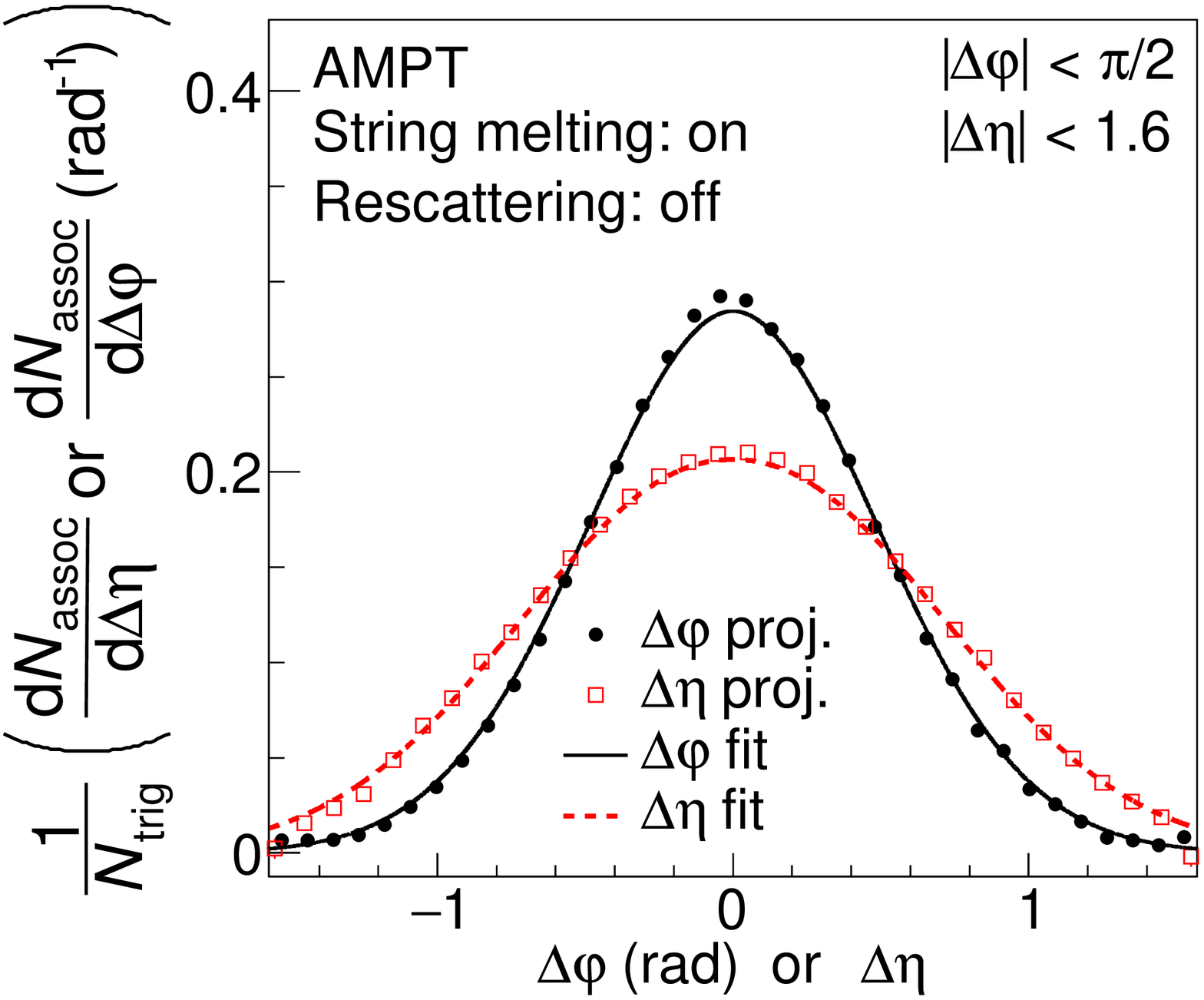}}
\subfigure[][]{\includegraphics[width=0.32\textwidth]{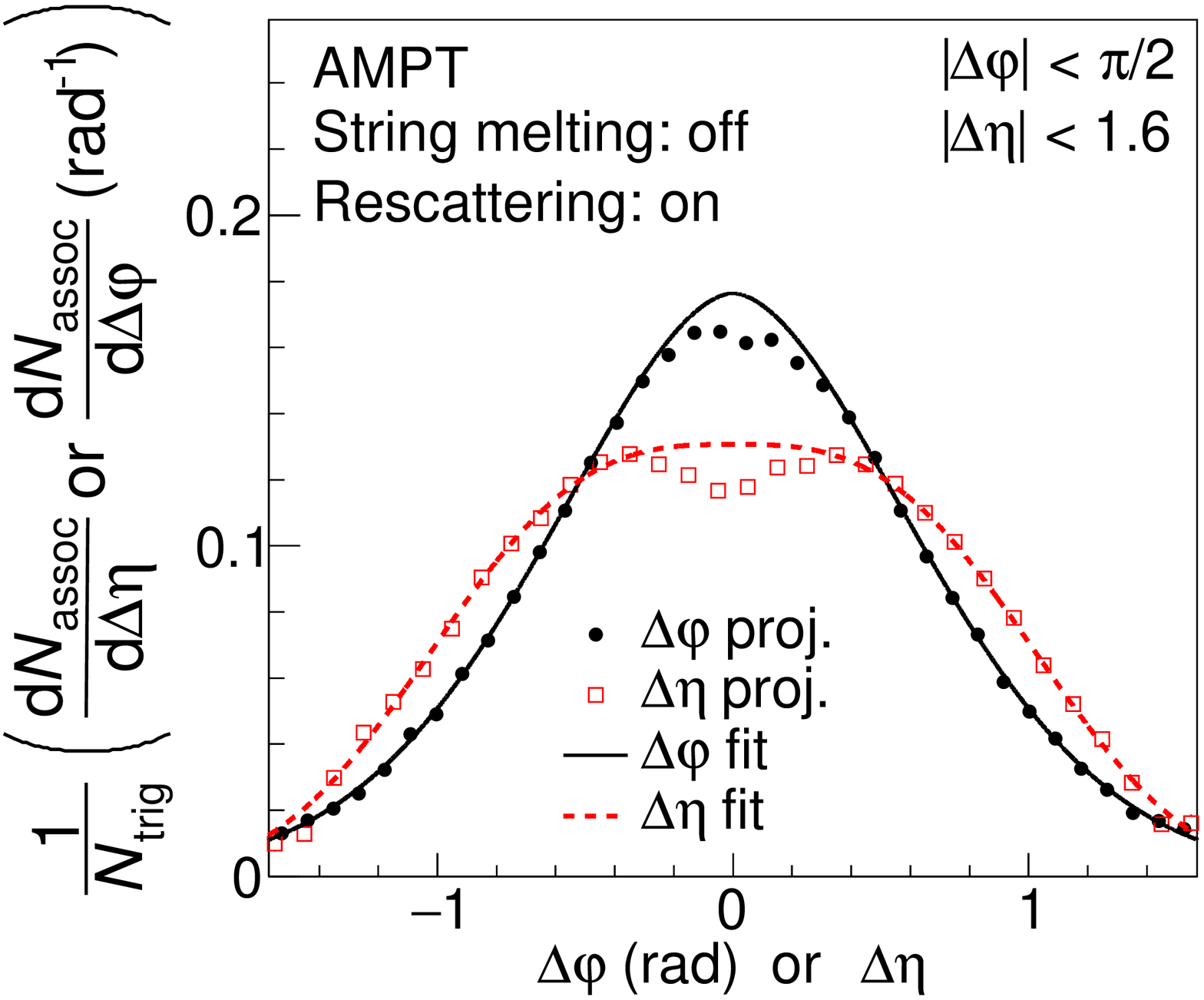}}
\caption{\label{fig:ampt_pertriggeryields}
 Upper row:  associated yield per trigger particle as a function of $\Dphi$ and $\Deta$ in AMPT (generator level) for the 10\% most central events.  Lower row:  projections to the $\Dphi$ and $\Deta$ axis. The bin shown is $1<\ptt<$~\unit[2]{\gevc} and $1<\pta<$~\unit[2]{\gevc}. Three different AMPT settings are shown. Left panels: string melting and hadronic rescattering active; middle panels: only string melting active; right panels: only hadronic rescattering active. As in Fig.~\ref{fig:pertriggeryields}, the combinatorial and flow background has been subtracted using the fit function.}
\end{figure*}

Figure~\ref{fig:ampt_pertriggeryields} presents the per-trigger yield and their projections to the $\Dphi$ and $\Deta$ axes for the AMPT simulations in the same $\pt$ and centrality bin as the top panel of Fig.~\ref{fig:pertriggeryields}. The AMPT simulations with hadronic rescattering show a depletion regardless of the string melting setting.

\begin{figure}[t!]
\centering
\includegraphics[width=0.49\textwidth]{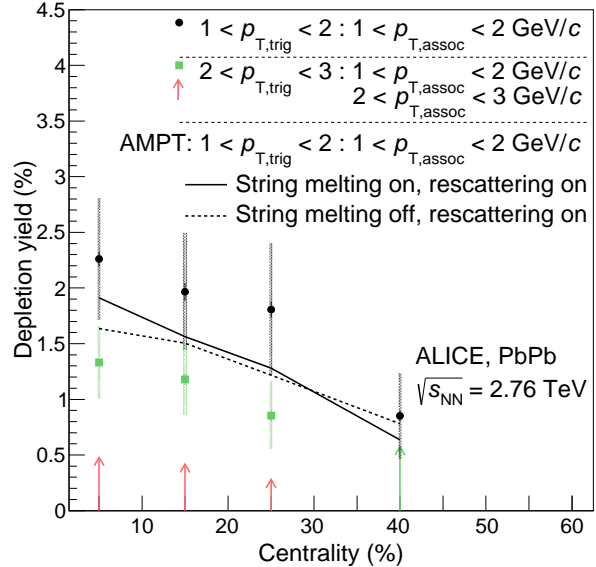}
\caption{\label{fig:depletionquantification}
Missing yield in the depletion region relative to the overall peak yield extracted from the fit. Lines indicate statistical uncertainties (mostly smaller than the marker size), while boxes (only for data) denote systematic uncertainties. The arrows indicate the upper limit in case the uncertainty bands touch 0. The markers are placed at the centre of the centrality bins. For comparison, the non-zero values from two AMPT simulations are shown as lines.}
\end{figure}

In order to quantify this depletion, the difference between the fit (where the depletion region has been excluded, see above) and the per-trigger yield relative to the total peak yield for the $\pt$ bins is computed and this is referred to as depletion yield in the following. The region where effects are expected from the limited two-track reconstruction efficiency ($|\Dphi| < 0.04$ and $|\Deta| < 0.05$, which corresponds to 0.5--6\% of the integrated region) is excluded from this calculation. Fig.~\ref{fig:depletionquantification} presents the depletion yield as a function of centrality for the $\pt$ bins where it is different from 0.
It can be seen that (2.2$\pm$0.5)\% of the yield is missing in the lowest $\pt$ bin ($1<\ptt<$~\unit[2]{\gevc}, $1<\pta<$~\unit[2]{\gevc}) and in the 10\% most central events. This value decreases gradually with centrality and with $\pt$. No significant depletion  is observed for 50--80\% (30--80\%) centrality or pp collisions for the lowest (second lowest) $\pt$ range. For higher $\pt$ bins, no significant depletion is observed.

The depletion observed in the AMPT events is present only in the lowest $\pt$ bin, where its value is compatible with the data for both settings where hadronic rescattering is switched on. For larger $\pt$ bins and for the configuration without hadronic rescattering the depletion yield is consistent with 0 in AMPT.

\subsection{Interpretation and relation to radial and elliptic flow}
A broadening of the near-side jet-like peak could originate from the modification of the jet fragmentation function in the medium. This is expected to manifest itself as a symmetric broadening in the $\Dphi$ and the $\Deta$ directions. The interaction of the penetrating jet with the flowing medium, could also result in a broadening of the peak, which could be of symmetric as well as of asymmetric shape. Therefore in order to investigate the relation of the observations and the strength of radial and anisotropic flow, Table~\ref{tab:v2beta} presents the radial-flow expansion velocity $\beta_{\rm T}$ and the elliptic flow coefficient $v_2\{2\}$ for the 10\% most central events from data (from \cite{Abelev:2013vea,Aamodt:2010pa}) and from the AMPT samples. The expansion velocity  $\beta_{\rm T}$ is extracted from a Blast-Wave fit to the $\pt$-spectra of $\pi$, K and p in the range of $0.5<\pt<$~\unit[1]{\gevc}, $0.2<\pt<$~\unit[1.5]{\gevc} and $0.3<\pt<$~\unit[2]{\gevc}, respectively, and in the rapidity range of $|y| < 0.5$. The fit describes the AMPT simulation with 10\% precision in the fitted range (see Ref.~\cite{Abelev:2013vea} for details on the fitting procedure). The $v_2\{2\}$ is extracted from two-particle correlations within $|\eta| < 0.8$ and \mbox{$0.2 < \pt < \unit[5]{\gevc}$} (see Ref.~\cite{Aamodt:2010pa} for details on the procedure).

\begin{table*}[bht!]
  \centering
  \begin{tabular}{l|c|c}
    \hline
    Sample & $\beta_{\rm T}$ & $v_2\{2\}$ \\
    \hline
    AMPT string melting and hadronic rescattering 	& 0.442 & 0.0412 $\pm$ 0.0002 \\
    AMPT string melting 				& 0.202 & 0.0389 $\pm$ 0.0002 \\
    AMPT hadronic rescattering 				& 0.540 & 0.0330 $\pm$ 0.0002 \\
    \hline
    Data 						& 0.649 $\pm$ 0.022 & 0.0364 $\pm$ 0.0003 \\
    \hline
  \end{tabular}
  \caption{\label{tab:v2beta}
    Blast-wave fit parameter $\beta_{\rm T}$ and elliptic flow coefficient $v_2\{2\}$ for 0--10\% centrality in Pb--Pb collisions at \snn\ = \unit[2.76]{TeV} in the considered AMPT samples and for comparison in the data (from \cite{Abelev:2013vea,Aamodt:2010pa}). Uncertainties are statistical for the MC samples and combined statistical and systematic ones for the data. The statistical uncertainties for the Blast-wave fits on AMPT are negligible.}
\end{table*}

The radial-flow expansion velocity $\beta_{\rm T}$ is larger when hadronic rescattering is active and largest if in addition string melting is switched off, while the configuration without hadronic rescattering results in a low $\beta_{\rm T}$. The value found in the data is about 20\% larger than the highest one in the AMPT simulations.
The elliptic flow coefficient $v_2\{2\}$ is better described by AMPT. Closest are the configurations with either string melting or hadronic rescattering (about 7\% discrepancy), while the configuration with both processes simultaneously overestimates the $v_2\{2\}$.
The differences between the different AMPT configurations are much smaller for the elliptic flow than for the radial flow.

The depletion discussed in the previous section occurs in the two AMPT configurations where the $\beta_{\rm T}$ is large, while the configuration without the depletion has the smallest $\beta_{\rm T}$.
The coefficient $v_2\{2\}$ has significantly different values in the two configurations with depletion, and the relative increase of the peak width (Fig.~\ref{fig:widthratios}) is best described by the AMPT configuration with the largest $\beta_{\rm T}$.
These studies suggest that the depletion is more likely accompanied by radial flow than by  elliptic flow.

Ref.~\cite{Ma:2008nd} studied partonic pseudorapidity distributions at different evolution times in AMPT. The authors show that the longitudinal broadening is driven by large values of longitudinal flow. In a picture where expansion is driven by pressure gradients, strong radial expansion can be accompanied by large longitudinal expansion. In conclusion, in AMPT, the observed phenomena are accompanied by large values of radial and longitudinal flow.

\subsection{Comparison to other experiments}

The STAR collaboration has studied near-side peak shapes at $\snn = \unit[62.4]{GeV}$ and $\snn = \unit[200]{GeV}$ in d+Au, Cu--Cu and Au--Au collisions~\cite{Agakishiev:2011st}. 
Apart from
the peak width quantification,  done separately in the $\Deta$ and $\Dphi$ direction with one-dimensional Gaussian functions after the subtraction of the background,
the analysis method is compatible to the one presented in this analysis.
In the studies presented in this paper, it was found that the peak widths with one two-dimensional Gaussian lead generally to smaller values than with the generalized Gaussian, and the fit quality is not optimal for the large statistics collected at the LHC. However, despite the difference in centre-of-mass energy, the larger statistical uncertainties in the analysis reported by the STAR collaboration may have hidden the possibility that the generalized Gaussian is a better description of the near-side peak.

\begin{figure*}[t!]
  \centering
  \subfigure[][]{\includegraphics[width=0.49\textwidth]{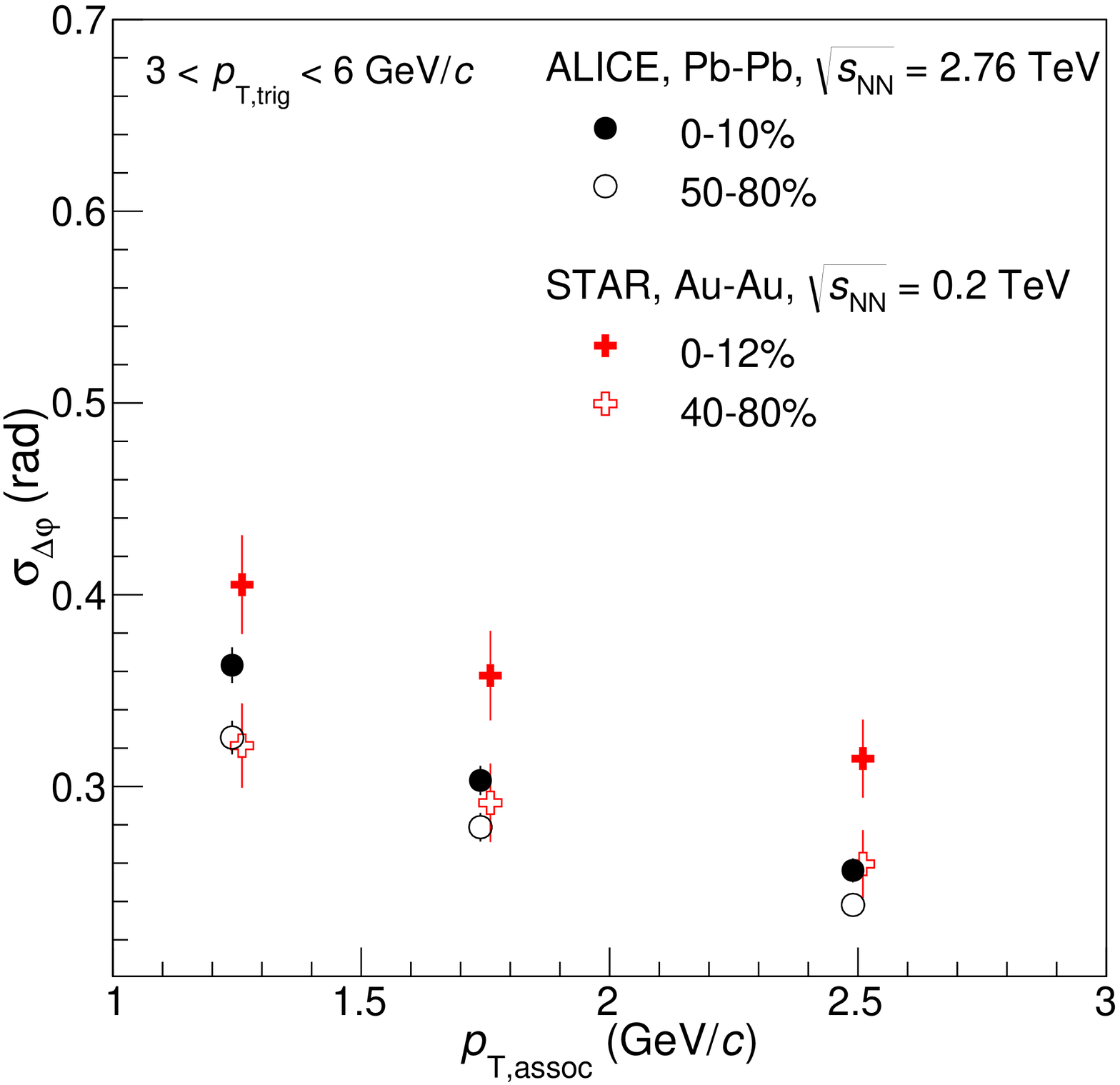}}
  \hfill
  \subfigure[][]{\includegraphics[width=0.49\textwidth]{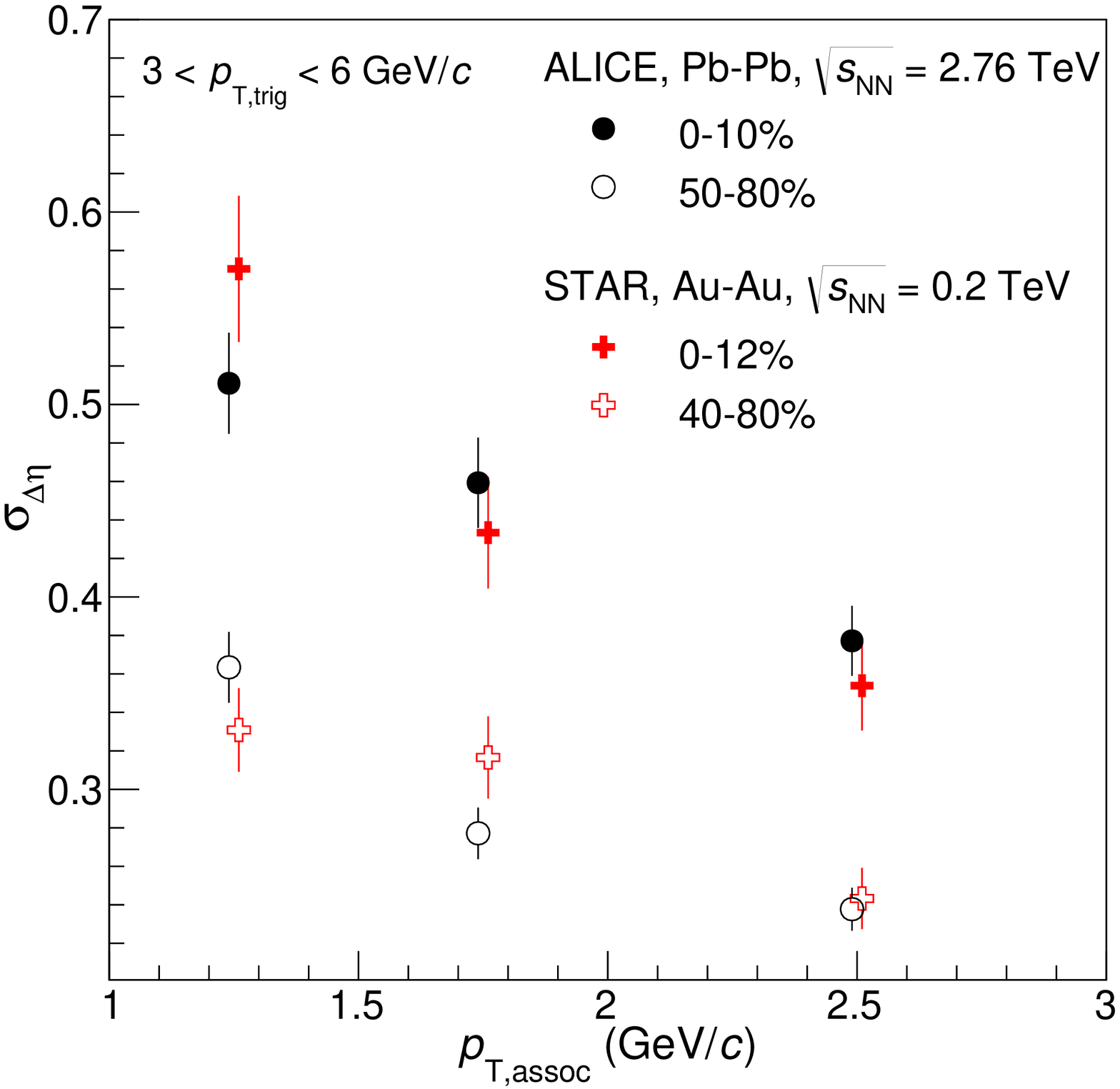}}
  \caption{\label{fig:star_comparison}
  Comparison of the shape parameters $\sigma_{\Dphi}$ (left panel) and $\sigma_{\Deta}$ (right panel) to results from the STAR collaboration in Au--Au collisions at $\snn = \unit[200]{GeV}$~\cite{Agakishiev:2011st}.
   The markers are placed at the centre of the $p_{\rm T,assoc}$ bins, slightly displaced for visibility, and the uncertainties are statistical and systematic ones added in quadrature.
}
\end{figure*}

The STAR collaboration also observed a broadening of the peak widths from peripheral to central collisions at $\snn = \unit[200]{GeV}$ in Au--Au collisions. The effect is most significant for $2 < \pta < \unit[4]{\gevc}$. In the $\Dphi$ direction, the width increases by 25--30\% depending on the $\pt$, and in the $\Deta$ direction, the increase is about 40--60\%. This effect is comparable to the observations presented in this paper.
Figure~\ref{fig:star_comparison} compares the results in the centrality bins which have the largest overlap between the two experiments. 
Agreement between the STAR results and this work is observed, within statistical uncertainties, in all overlapping momentum and centrality bins, with the exception of
central collisions in the $\Dphi$ direction, where a two-sigma difference can be seen.

\section{Summary}
\label{sec:summary}

We have presented a detailed characterization of the flow-subtracted near-side peak in two-particle correlations in Pb--Pb collisions at $\snn=$~\unit[2.76]{TeV} together with a measurement in pp collisions at the same energy. The near-side peak shows the characteristic $\pt$ dependence, where the width decreases with increasing $\pt$. In addition, in Pb--Pb collisions a centrality dependence is observed: the peak gets wider from peripheral to central collisions. This effect is significantly stronger for the $\Deta$ direction than for the $\Dphi$ direction leading to an asymmetric peak shape in central collisions, observed over a wide $\pt$ range.
Additionally, at low $\pt$, an unexpected depletion develops from peripheral to central collisions.

AMPT simulations show that both the asymmetric broadening and the depletion are also present when hadronic rescattering is included. The AMPT configuration with hadronic rescattering and without string melting reproduces quantitatively the relative peak broadening as well as the size of the depletion.
The extraction of the radial-flow expansion velocity suggests that the stronger the radial flow, the stronger the observed effects are. In addition, earlier theoretical and phenomenological work connected the longitudinal broadening of the near-side jet-like peak to strong longitudinal flow in AMPT, as well as to an interplay of partons traversing the longitudinally expanding medium.
Thus a possible scenario is that the presented observations are caused by the interplay of the jet with the collective expansion.

\ifpreprint
\iffull
\newenvironment{acknowledgement}{\relax}{\relax}
\begin{acknowledgement}
\section*{Acknowledgements}

The ALICE Collaboration would like to thank all its engineers and technicians for their invaluable contributions to the construction of the experiment and the CERN accelerator teams for the outstanding performance of the LHC complex.
The ALICE Collaboration gratefully acknowledges the resources and support provided by all Grid centres and the Worldwide LHC Computing Grid (WLCG) collaboration.
The ALICE Collaboration acknowledges the following funding agencies for their support in building and running the ALICE detector:
A. I. Alikhanyan National Science Laboratory (Yerevan Physics Institute) Foundation (ANSL), State Committee of Science and World Federation of Scientists (WFS), Armenia;
Austrian Academy of Sciences and Nationalstiftung f\"{u}r Forschung, Technologie und Entwicklung, Austria;
, Conselho Nacional de Desenvolvimento Cient\'{\i}fico e Tecnol\'{o}gico (CNPq), Financiadora de Estudos e Projetos (Finep) and Funda\c{c}\~{a}o de Amparo \`{a} Pesquisa do Estado de S\~{a}o Paulo (FAPESP), Brazil;
Ministry of Education of China (MOE of China), Ministry of Science \& Technology of China (MOST of China) and National Natural Science Foundation of China (NSFC), China;
Ministry of Science, Education and Sport and Croatian Science Foundation, Croatia;
Centro de Investigaciones Energ\'{e}ticas, Medioambientales y Tecnol\'{o}gicas (CIEMAT), Cuba;
Ministry of Education, Youth and Sports of the Czech Republic, Czech Republic;
Danish National Research Foundation (DNRF), The Carlsberg Foundation and The Danish Council for Independent Research | Natural Sciences, Denmark;
Helsinki Institute of Physics (HIP), Finland;
Commissariat \`{a} l'Energie Atomique (CEA) and Institut National de Physique Nucl\'{e}aire et de Physique des Particules (IN2P3) and Centre National de la Recherche Scientifique (CNRS), France;
Bundesministerium f\"{u}r Bildung, Wissenschaft, Forschung und Technologie (BMBF) and GSI Helmholtzzentrum f\"{u}r Schwerionenforschung GmbH, Germany;
Ministry of Education, Research and Religious Affairs, Greece;
National Research, Development and Innovation Office, Hungary;
Department of Atomic Energy Government of India (DAE), India;
Indonesian Institute of Science, Indonesia;
Centro Fermi - Museo Storico della Fisica e Centro Studi e Ricerche Enrico Fermi and Istituto Nazionale di Fisica Nucleare (INFN), Italy;
Institute for Innovative Science and Technology , Nagasaki Institute of Applied Science (IIST), Japan Society for the Promotion of Science (JSPS) KAKENHI and Japanese Ministry of Education, Culture, Sports, Science and Technology (MEXT), Japan;
Consejo Nacional de Ciencia (CONACYT) y Tecnolog\'{i}a, through Fondo de Cooperaci\'{o}n Internacional en Ciencia y Tecnolog\'{i}a (FONCICYT) and Direcci\'{o}n General de Asuntos del Personal Academico (DGAPA), Mexico;
Nationaal instituut voor subatomaire fysica (Nikhef), Netherlands;
The Research Council of Norway, Norway;
Commission on Science and Technology for Sustainable Development in the South (COMSATS), Pakistan;
Pontificia Universidad Cat\'{o}lica del Per\'{u}, Peru;
Ministry of Science and Higher Education and National Science Centre, Poland;
Ministry of Education and Scientific Research, Institute of Atomic Physics and Romanian National Agency for Science, Technology and Innovation, Romania;
Joint Institute for Nuclear Research (JINR), Ministry of Education and Science of the Russian Federation and National Research Centre Kurchatov Institute, Russia;
Ministry of Education, Science, Research and Sport of the Slovak Republic, Slovakia;
National Research Foundation of South Africa, South Africa;
Korea Institute of Science and Technology Information and National Research Foundation of Korea (NRF), South Korea;
Centro de Investigaciones Energ\'{e}ticas, Medioambientales y Tecnol\'{o}gicas (CIEMAT) and Ministerio de Ciencia e Innovacion, Spain;
Knut \& Alice Wallenberg Foundation (KAW) and Swedish Research Council (VR), Sweden;
European Organization for Nuclear Research, Switzerland;
National Science and Technology Development Agency (NSDTA), Office of the Higher Education Commission under NRU project of Thailand and Suranaree University of Technology (SUT), Thailand;
Turkish Atomic Energy Agency (TAEK), Turkey;
National Academy of  Sciences of Ukraine, Ukraine;
Science and Technology Facilities Council (STFC), United Kingdom;
National Science Foundation of the United States of America (NSF) and United States Department of Energy, Office of Nuclear Physics (DOE NP), United States.    
\end{acknowledgement}
\fi
\ifbibtex
\bibliographystyle{utphys}
\bibliography{biblio}{}
\else
\input{references_cern.tex}
\fi
\iffull
\newpage
\appendix
\section{The ALICE Collaboration}
\label{app:collab}



\begingroup
\small
\begin{flushleft}
J.~Adam$^\textrm{\scriptsize 39}$,
D.~Adamov\'{a}$^\textrm{\scriptsize 86}$,
M.M.~Aggarwal$^\textrm{\scriptsize 90}$,
G.~Aglieri Rinella$^\textrm{\scriptsize 35}$,
M.~Agnello$^\textrm{\scriptsize 113}$\textsuperscript{,}$^\textrm{\scriptsize 31}$,
N.~Agrawal$^\textrm{\scriptsize 48}$,
Z.~Ahammed$^\textrm{\scriptsize 137}$,
S.~Ahmad$^\textrm{\scriptsize 18}$,
S.U.~Ahn$^\textrm{\scriptsize 70}$,
S.~Aiola$^\textrm{\scriptsize 141}$
% ...
\renewcommand\labelenumi{\textsuperscript{\theenumi}~}

\section*{Affiliation notes}
\renewcommand\theenumi{\roman{enumi}}
\begin{Authlist}
\item \Adef{0}Deceased
\item \Adef{idp26411852}{Also at: Georgia State University, Atlanta, Georgia, United States}
\item \Adef{idp27127172}{Also at: Also at Department of Applied Physics, Aligarh Muslim University, Aligarh, India}
\item \Adef{idp27505124}{Also at: M.V. Lomonosov Moscow State University, D.V. Skobeltsyn Institute of Nuclear, Physics, Moscow, Russia}
\end{Authlist}

\section*{Collaboration Institutes}
\renewcommand\theenumi{\arabic{enumi}~}

$^{1}$A.I. Alikhanyan National Science Laboratory (Yerevan Physics Institute) Foundation, Yerevan, Armenia
\\
$^{2}$Benem\'{e}rita Universidad Aut\'{o}noma de Puebla, Puebla, Mexico
\\
$^{3}$Bogolyubov Institute for Theoretical Physics, Kiev, Ukraine
\\
$^{4}$Bose Institute, Department of Physics
and Centre for Astroparticle Physics and Space Science (CAPSS), Kolkata, India
\\
$^{5}$Budker Institute for Nuclear Physics, Novosibirsk, Russia
\\
$^{6}$California Polytechnic State University, San Luis Obispo, California, United States
\\
$^{7}$Central China Normal University, Wuhan, China
\\
$^{8}$Centre de Calcul de l'IN2P3, Villeurbanne, Lyon, France
\\
$^{9}$Centro de Aplicaciones Tecnol\'{o}gicas y Desarrollo Nuclear (CEADEN), Havana, Cuba
\\
$^{10}$Centro de Investigaciones Energ\'{e}ticas Medioambientales y Tecnol\'{o}gicas (CIEMAT), Madrid, Spain
\\
$^{11}$Centro de Investigaci\'{o}n y de Estudios Avanzados (CINVESTAV), Mexico City and M\'{e}rida, Mexico
\\
$^{12}$Centro Fermi - Museo Storico della Fisica e Centro Studi e Ricerche ``Enrico Fermi', Rome, Italy
\\
$^{13}$Chicago State University, Chicago, Illinois, United States
\\
$^{14}$China Institute of Atomic Energy, Beijing, China
\\
$^{15}$Commissariat \`{a} l'Energie Atomique, IRFU, Saclay, France
\\
$^{16}$COMSATS Institute of Information Technology (CIIT), Islamabad, Pakistan
\\
$^{17}$Departamento de F\'{\i}sica de Part\'{\i}culas and IGFAE, Universidad de Santiago de Compostela, Santiago de Compostela, Spain
\\
$^{18}$Department of Physics, Aligarh Muslim University, Aligarh, India
\\
$^{19}$Department of Physics, Ohio State University, Columbus, Ohio, United States
\\
$^{20}$Department of Physics, Sejong University, Seoul, South Korea
\\
$^{21}$Department of Physics, University of Oslo, Oslo, Norway
\\
$^{22}$Department of Physics and Technology, University of Bergen, Bergen, Norway
\\
$^{23}$Dipartimento di Fisica dell'Universit\`{a} 'La Sapienza'
and Sezione INFN, Rome, Italy
\\
$^{24}$Dipartimento di Fisica dell'Universit\`{a}
and Sezione INFN, Cagliari, Italy
\\
$^{25}$Dipartimento di Fisica dell'Universit\`{a}
and Sezione INFN, Trieste, Italy
\\
$^{26}$Dipartimento di Fisica dell'Universit\`{a}
and Sezione INFN, Turin, Italy
\\
$^{27}$Dipartimento di Fisica e Astronomia dell'Universit\`{a}
and Sezione INFN, Bologna, Italy
\\
$^{28}$Dipartimento di Fisica e Astronomia dell'Universit\`{a}
and Sezione INFN, Catania, Italy
\\
$^{29}$Dipartimento di Fisica e Astronomia dell'Universit\`{a}
and Sezione INFN, Padova, Italy
\\
$^{30}$Dipartimento di Fisica `E.R.~Caianiello' dell'Universit\`{a}
and Gruppo Collegato INFN, Salerno, Italy
\\
$^{31}$Dipartimento DISAT del Politecnico and Sezione INFN, Turin, Italy
\\
$^{32}$Dipartimento di Scienze e Innovazione Tecnologica dell'Universit\`{a} del Piemonte Orientale and INFN Sezione di Torino, Alessandria, Italy
\\
$^{33}$Dipartimento Interateneo di Fisica `M.~Merlin'
and Sezione INFN, Bari, Italy
\\
$^{34}$Division of Experimental High Energy Physics, University of Lund, Lund, Sweden
\\
$^{35}$European Organization for Nuclear Research (CERN), Geneva, Switzerland
\\
$^{36}$Excellence Cluster Universe, Technische Universit\"{a}t M\"{u}nchen, Munich, Germany
\\
$^{37}$Faculty of Engineering, Bergen University College, Bergen, Norway
\\
$^{38}$Faculty of Mathematics, Physics and Informatics, Comenius University, Bratislava, Slovakia
\\
$^{39}$Faculty of Nuclear Sciences and Physical Engineering, Czech Technical University in Prague, Prague, Czech Republic
\\
$^{40}$Faculty of Science, P.J.~\v{S}af\'{a}rik University, Ko\v{s}ice, Slovakia
\\
$^{41}$Faculty of Technology, Buskerud and Vestfold University College, Tonsberg, Norway
\\
$^{42}$Frankfurt Institute for Advanced Studies, Johann Wolfgang Goethe-Universit\"{a}t Frankfurt, Frankfurt, Germany
\\
$^{43}$Gangneung-Wonju National University, Gangneung, South Korea
\\
$^{44}$Gauhati University, Department of Physics, Guwahati, India
\\
$^{45}$Helmholtz-Institut f\"{u}r Strahlen- und Kernphysik, Rheinische Friedrich-Wilhelms-Universit\"{a}t Bonn, Bonn, Germany
\\
$^{46}$Helsinki Institute of Physics (HIP), Helsinki, Finland
\\
$^{47}$Hiroshima University, Hiroshima, Japan
\\
$^{48}$Indian Institute of Technology Bombay (IIT), Mumbai, India
\\
$^{49}$Indian Institute of Technology Indore, Indore, India
\\
$^{50}$Indonesian Institute of Sciences, Jakarta, Indonesia
\\
$^{51}$Inha University, Incheon, South Korea
\\
$^{52}$Institut de Physique Nucl\'eaire d'Orsay (IPNO), Universit\'e Paris-Sud, CNRS-IN2P3, Orsay, France
\\
$^{53}$Institute for Nuclear Research, Academy of Sciences, Moscow, Russia
\\
$^{54}$Institute for Subatomic Physics of Utrecht University, Utrecht, Netherlands
\\
$^{55}$Institute for Theoretical and Experimental Physics, Moscow, Russia
\\
$^{56}$Institute of Experimental Physics, Slovak Academy of Sciences, Ko\v{s}ice, Slovakia
\\
$^{57}$Institute of Physics, Academy of Sciences of the Czech Republic, Prague, Czech Republic
\\
$^{58}$Institute of Physics, Bhubaneswar, India
\\
$^{59}$Institute of Space Science (ISS), Bucharest, Romania
\\
$^{60}$Institut f\"{u}r Informatik, Johann Wolfgang Goethe-Universit\"{a}t Frankfurt, Frankfurt, Germany
\\
$^{61}$Institut f\"{u}r Kernphysik, Johann Wolfgang Goethe-Universit\"{a}t Frankfurt, Frankfurt, Germany
\\
$^{62}$Institut f\"{u}r Kernphysik, Westf\"{a}lische Wilhelms-Universit\"{a}t M\"{u}nster, M\"{u}nster, Germany
\\
$^{63}$Instituto de Ciencias Nucleares, Universidad Nacional Aut\'{o}noma de M\'{e}xico, Mexico City, Mexico
\\
$^{64}$Instituto de F\'{i}sica, Universidade Federal do Rio Grande do Sul (UFRGS), Porto Alegre, Brazil
\\
$^{65}$Instituto de F\'{\i}sica, Universidad Nacional Aut\'{o}noma de M\'{e}xico, Mexico City, Mexico
\\
$^{66}$Institut Pluridisciplinaire Hubert Curien (IPHC), Universit\'{e} de Strasbourg, CNRS-IN2P3, Strasbourg, France, Strasbourg, France
\\
$^{67}$iThemba LABS, National Research Foundation, Somerset West, South Africa
\\
$^{68}$Joint Institute for Nuclear Research (JINR), Dubna, Russia
\\
$^{69}$Konkuk University, Seoul, South Korea
\\
$^{70}$Korea Institute of Science and Technology Information, Daejeon, South Korea
\\
$^{71}$KTO Karatay University, Konya, Turkey
\\
$^{72}$Laboratoire de Physique Corpusculaire (LPC), Clermont Universit\'{e}, Universit\'{e} Blaise Pascal, CNRS--IN2P3, Clermont-Ferrand, France
\\
$^{73}$Laboratoire de Physique Subatomique et de Cosmologie, Universit\'{e} Grenoble-Alpes, CNRS-IN2P3, Grenoble, France
\\
$^{74}$Laboratori Nazionali di Frascati, INFN, Frascati, Italy
\\
$^{75}$Laboratori Nazionali di Legnaro, INFN, Legnaro, Italy
\\
$^{76}$Lawrence Berkeley National Laboratory, Berkeley, California, United States
\\
$^{77}$Moscow Engineering Physics Institute, Moscow, Russia
\\
$^{78}$Nagasaki Institute of Applied Science, Nagasaki, Japan
\\
$^{79}$National Centre for Nuclear Studies, Warsaw, Poland
\\
$^{80}$National Institute for Physics and Nuclear Engineering, Bucharest, Romania
\\
$^{81}$National Institute of Science Education and Research, Bhubaneswar, India
\\
$^{82}$National Research Centre Kurchatov Institute, Moscow, Russia
\\
$^{83}$Niels Bohr Institute, University of Copenhagen, Copenhagen, Denmark
\\
$^{84}$Nikhef, Nationaal instituut voor subatomaire fysica, Amsterdam, Netherlands
\\
$^{85}$Nuclear Physics Group, STFC Daresbury Laboratory, Daresbury, United Kingdom
\\
$^{86}$Nuclear Physics Institute, Academy of Sciences of the Czech Republic, \v{R}e\v{z} u Prahy, Czech Republic, \v{R}e\v{z} u Prahy, Czech Republic
\\
$^{87}$Oak Ridge National Laboratory, Oak Ridge, Tennessee, United States
\\
$^{88}$Petersburg Nuclear Physics Institute, Gatchina, Russia
\\
$^{89}$Physics Department, Creighton University, Omaha, Nebraska, United States
\\
$^{90}$Physics Department, Panjab University, Chandigarh, India
\\
$^{91}$Physics Department, University of Athens, Athens, Greece
\\
$^{92}$Physics Department, University of Cape Town, Cape Town, South Africa
\\
$^{93}$Physics Department, University of Jammu, Jammu, India
\\
$^{94}$Physics Department, University of Rajasthan, Jaipur, India
\\
$^{95}$Physikalisches Institut, Eberhard Karls Universit\"{a}t T\"{u}bingen, T\"{u}bingen, Germany
\\
$^{96}$Physikalisches Institut, Ruprecht-Karls-Universit\"{a}t Heidelberg, Heidelberg, Germany
\\
$^{97}$Physik Department, Technische Universit\"{a}t M\"{u}nchen, Munich, Germany
\\
$^{98}$Purdue University, West Lafayette, Indiana, United States
\\
$^{99}$Pusan National University, Pusan, South Korea
\\
$^{100}$Research Division and ExtreMe Matter Institute EMMI, GSI Helmholtzzentrum f\"ur Schwerionenforschung, Darmstadt, Germany
\\
$^{101}$Rudjer Bo\v{s}kovi\'{c} Institute, Zagreb, Croatia
\\
$^{102}$Russian Federal Nuclear Center (VNIIEF), Sarov, Russia
\\
$^{103}$Saha Institute of Nuclear Physics, Kolkata, India
\\
$^{104}$School of Physics and Astronomy, University of Birmingham, Birmingham, United Kingdom
\\
$^{105}$Secci\'{o}n F\'{\i}sica, Departamento de Ciencias, Pontificia Universidad Cat\'{o}lica del Per\'{u}, Lima, Peru
\\
$^{106}$Sezione INFN, Bari, Italy
\\
$^{107}$Sezione INFN, Bologna, Italy
\\
$^{108}$Sezione INFN, Cagliari, Italy
\\
$^{109}$Sezione INFN, Catania, Italy
\\
$^{110}$Sezione INFN, Padova, Italy
\\
$^{111}$Sezione INFN, Rome, Italy
\\
$^{112}$Sezione INFN, Trieste, Italy
\\
$^{113}$Sezione INFN, Turin, Italy
\\
$^{114}$SSC IHEP of NRC Kurchatov institute, Protvino, Russia
\\
$^{115}$Stefan Meyer Institut f\"{u}r Subatomare Physik (SMI), Vienna, Austria
\\
$^{116}$SUBATECH, Ecole des Mines de Nantes, Universit\'{e} de Nantes, CNRS-IN2P3, Nantes, France
\\
$^{117}$Suranaree University of Technology, Nakhon Ratchasima, Thailand
\\
$^{118}$Technical University of Ko\v{s}ice, Ko\v{s}ice, Slovakia
\\
$^{119}$Technical University of Split FESB, Split, Croatia
\\
$^{120}$The Henryk Niewodniczanski Institute of Nuclear Physics, Polish Academy of Sciences, Cracow, Poland
\\
$^{121}$The University of Texas at Austin, Physics Department, Austin, Texas, United States
\\
$^{122}$Universidad Aut\'{o}noma de Sinaloa, Culiac\'{a}n, Mexico
\\
$^{123}$Universidade de S\~{a}o Paulo (USP), S\~{a}o Paulo, Brazil
\\
$^{124}$Universidade Estadual de Campinas (UNICAMP), Campinas, Brazil
\\
$^{125}$Universidade Federal do ABC, Santo Andre, Brazil
\\
$^{126}$University of Houston, Houston, Texas, United States
\\
$^{127}$University of Jyv\"{a}skyl\"{a}, Jyv\"{a}skyl\"{a}, Finland
\\
$^{128}$University of Liverpool, Liverpool, United Kingdom
\\
$^{129}$University of Tennessee, Knoxville, Tennessee, United States
\\
$^{130}$University of the Witwatersrand, Johannesburg, South Africa
\\
$^{131}$University of Tokyo, Tokyo, Japan
\\
$^{132}$University of Tsukuba, Tsukuba, Japan
\\
$^{133}$University of Zagreb, Zagreb, Croatia
\\
$^{134}$Universit\'{e} de Lyon, Universit\'{e} Lyon 1, CNRS/IN2P3, IPN-Lyon, Villeurbanne, Lyon, France
\\
$^{135}$Universit\`{a} di Brescia, Brescia, Italy
\\
$^{136}$V.~Fock Institute for Physics, St. Petersburg State University, St. Petersburg, Russia
\\
$^{137}$Variable Energy Cyclotron Centre, Kolkata, India
\\
$^{138}$Warsaw University of Technology, Warsaw, Poland
\\
$^{139}$Wayne State University, Detroit, Michigan, United States
\\
$^{140}$Wigner Research Centre for Physics, Hungarian Academy of Sciences, Budapest, Hungary
\\
$^{141}$Yale University, New Haven, Connecticut, United States
\\
$^{142}$Yonsei University, Seoul, South Korea
\\
$^{143}$Zentrum f\"{u}r Technologietransfer und Telekommunikation (ZTT), Fachhochschule Worms, Worms, Germany
\endgroup

\fi
\else 
\iffull
\\

\providecommand{\href}[2]{#2}\begingroup\raggedright\endgroup

\else
\ifbibtex
\bibliographystyle{apsrev4-1}
\bibliography{biblio}{}
\else

\fi
\fi
\fi
\end{document}